\documentclass[lettersize, journal]{IEEEtran}
\usepackage{amsmath, amssymb, amsfonts}
\usepackage[compress]{cite}
\usepackage{multirow}
\usepackage{booktabs}
\usepackage{enumitem}
\usepackage{graphicx}

\def\BibTeX{{\rm B\kern-.05em{\sc i\kern-.025em b}\kern-.08em
		T\kern-.1667em\lower.7ex\hbox{E}\kern-.125emX}}

\usepackage[labelformat=simple]{subcaption}
\newcommand{\refsec}[1]{Section~\ref{#1}}
\newcommand{\reffig}[1]{Fig.~\ref{#1}}
\newcommand{\reftab}[1]{Table~\ref{#1}}
\newcommand{\refequ}[1]{Eqn.~\eqref{#1}}
\newcommand{\tabincell}[2]{\begin{tabular}{@{}#1@{}}#2\end{tabular}}

\newcommand{\etal}{\emph{et al}. }
\newcommand{\ie}{\emph{i}.\emph{e}.}
\newcommand{\eg}{\emph{e}.\emph{g}.}

\begin{document}
	\title{GCNs-Net: A Graph Convolutional Neural Network Approach for Decoding Time-resolved EEG Motor Imagery Signals}
	\author{Yimin Hou *, Shuyue Jia *, \IEEEmembership{Student Member, IEEE}, Xiangmin Lun, Ziqian Hao, Yan Shi, Yang Li, \IEEEmembership{Senior Member, IEEE}, Rui Zeng, and Jinglei Lv
		
		\IEEEcompsocitemizethanks{
			\IEEEcompsocthanksitem * indicates co-first authorship.
			\IEEEcompsocthanksitem Yimin Hou, Xiangmin Lun, and Yan Shi are with the School of Automation Engineering, Northeast Electric Power University, Jilin 132011, China (e-mail: ymh7821@163.com; xm\_lun77@163.com; shiyan@neepu.edu.cn).
			\IEEEcompsocthanksitem This work was done while Shuyue Jia was with the School of Automation Engineering, Northeast Electric Power University, Jilin 132011, China. He is now with the Department of Computer Science, City University of Hong Kong, Hong Kong, China (e-mail: shuyuej@ieee.org).
			\IEEEcompsocthanksitem Ziqian Hao is with the Jinan Vocational College of Nursing, Jinan 250102, China (e-mail: ziqhao@163.com).
			\IEEEcompsocthanksitem Yang Li is with the School of Electrical Engineering, Northeast Electric Power University, Jilin 132011, China (e-mail: liyang@neepu.edu.cn).
			\IEEEcompsocthanksitem Rui Zeng and Jinglei Lv are with School of Biomedical Engineering \& Brain and Mind Center, University of Sydney, Sydney 2006, NSW, Australia (e-mail: rui.zeng@sydney.edu.au; jinglei.lv@sydney.edu.au).
		}
	}
	
	\markboth{Journal of \LaTeX\ Class Files,~Vol.~18, No.~9, September~2020}
	{How to Use the IEEEtran \LaTeX \ Templates}
	
	\IEEEtitleabstractindextext{
		\begin{abstract}
			Towards developing effective and efficient brain-computer interface (BCI) systems, precise decoding of brain activity measured by electroencephalogram (EEG), is highly demanded. Traditional works classify EEG signals without considering the topological relationship among electrodes. However, neuroscience research has increasingly emphasized network patterns of brain dynamics. Thus, the Euclidean structure of electrodes might not adequately reflect the interaction between signals. To fill the gap, a novel deep learning framework based on the graph convolutional neural networks (GCNs) is presented to enhance the decoding performance of raw EEG signals during different types of motor imagery (MI) tasks while cooperating with the functional topological relationship of electrodes. Based on the absolute Pearson’s matrix of overall signals, the graph Laplacian of EEG electrodes is built up. The GCNs-Net constructed by graph convolutional layers learns the generalized features. The followed pooling layers reduce dimensionality, and the fully-connected softmax layer derives the final prediction. The introduced approach has been shown to converge for both personalized and group-wise predictions. It has achieved the highest averaged accuracy, 93.06\% and 88.57\% (PhysioNet Dataset), 96.24\% and 80.89\% (High Gamma Dataset), at the subject and group level, respectively, compared with existing studies, which suggests adaptability and robustness to individual variability. Moreover, the performance is stably reproducible among repetitive experiments for cross-validation. The excellent performance of our method has shown that it is an important step towards better BCI approaches. To conclude, the GCNs-Net filters EEG signals based on the functional topological relationship, which manages to decode relevant features for brain motor imagery. A deep learning library for EEG tasks classification including the code for this study is open source at https://github.com/SuperBruceJia/EEG-DL for scientific research.
		\end{abstract}
		
	}
	
	\maketitle
	\IEEEdisplaynontitleabstractindextext
	
	\section{Introduction}\label{Introduction}
	\IEEEPARstart{R}{ecently}, the brain-computer interface (BCI) has become one of the hottest research topics for broad applications in the field of therapeutic and medical engineering~\cite{santhanam2006high}. It refers to the establishment of an innovative technology that exchanges information directly between the brain and the surroundings, which does not rely on traditional methods such as human muscle tissue or peripheral nerves. BCI systems decode brain activity patterns to manipulate assistant devices, such as wheelchairs and artificial limbs~\cite{silver2005topographic}. Electroencephalogram (EEG) is extensively applied because of its high temporal resolution, noninvasiveness, and portability. The principle of EEG is to record spontaneous, event-related, and stimulus-evoked electrical signals of the brain on time scales, which reveals variations for different brain activities~\cite{schomer2012niedermeyer}. EEG decodes discriminable brain patterns while carrying out different types of actual movement or imagery~\cite{hubbard2019eeg, mcfarland2000mu}. Motor imagery (MI) based EEG mentally simulates multiple motor motions, such as imagining hand or foot movements. Controlling machines via only the MI without physical movements of the body is one of the elemental jobs in BCI~\cite{lee2018convolution}. To realize such BCI systems, accurate classification of MI brain activity is of great essence. Although previous studies have shown promising performances, there is still space to improve the classification accuracy towards building effective and efficient BCI applications. For instance, the adaptability and robustness to individual variability remain among the challenges of setting up an EEG MI-based wheelchair. Traditional approaches do not consider the topological relationship of electrodes while decoding EEG signals. However, a growing number of neuroscience research has emphasized brain network dynamics~\cite{bassett2017network, lv2014holistic, lv2017n}. Thus, the interaction between signals might not be adequately reflected and represented via the Euclidean structure of EEG electrodes. To address the concern, the Graph Convolutional Neural Networks (GCNs) are introduced to decode EEG signals, promoting the classification performance by cooperating with the functional topological relationship of EEG electrodes and implementing the Convolutional Neural Networks (CNNs) on graphs.
	
	\section{Literature Survey}\label{Literature Survey}
	Traditional works manually designed features from EEG signals,~\eg, via the analytic intrinsic mode functions or wavelet transform, and then employed machine learning-based approaches to classify features~\cite{taran2018features, yu2015enhanced, edelman2015eeg, wu2008classifying, he2019transfer}. Recently, deep learning (DL) has achieved superhuman performances across multiple domains~\cite{lecun2015deep, 9772641, ZHANG2022113824}. The DL-based methods learned the underlying features from signals, which alleviated the need for hands-on feature engineering. The CNNs have been broadly employed to classify the Euclidean-structured signals on account of their ability to learn informative features via the local receptive fields. The CNNs-based approaches~\cite{lawhern2018eegnet, faust2018deep, hou2019novel, amin2019deep, dose2018end, chaudhary2019convolutional, schirrmeister2017deep, ortiz2019new, li2019densely, zhang2019novel, alazrai2019deep, tang2020conditional} were implemented to address the challenge of EEG task classification. In~\cite{hou2019novel}, Hou~\etal introduced an innovative approach by combining the Scout EEG Source Imaging (ESI) and CNNs to decode EEG tasks, which achieved competitive results, 94.5\% maximum accuracy for 10 subjects, and 92.5\% for 14 subjects, on the PhysioNet Dataset~\cite{goldberger2000physiobank}. Zhang~\etal\cite{zhang2019making} presented a cascade convolutional recurrent neural network, and obtained 98.31\% averaged accuracy on the PhysioNet Dataset. Reference~\cite{dose2018end} applied one-dimensional convolutional filters to learn the temporal and spatial features, and it reached 80.38\%, 69.82\%, and 58.58\% accuracy on the PhysioNet Dataset with two, three, and four MI tasks. References~\cite{schirrmeister2017deep, amin2019deep, li2019densely, zhang2019novel} utilized variants of CNNs to decode EEG signals from the BCI Competition IV-2a Dataset~\cite{brunner2008bci}, and achieved 73.70\%, 75.70\%, 79.90\%, and 83.00\% accuracy, respectively. References~\cite{alazrai2019deep, tang2020conditional, amin2019deep} obtained 92.50\%, 93.70\%, and 95.4\% accuracy at the subject level on the High Gamma Dataset~\cite{tangermann2012review}. Although the performance of the above CNNs-based models was encouraging, there was still space to promote the classification accuracy to build a robust and reliable BCI system. The reasons why we applied the GCNs were as follows. The traditional CNNs cannot directly process the non-Euclidean structured data because the discrete convolutions cannot keep translation invariance on the non-Euclidean signals. However, the GCNs can directly extract features from the non-Euclidean data and process the graph-structured signals since the GCNs consider the relationship properties (\eg, correlations) between nodes~\cite{hou2022deep, defferrard2016convolutional}. Through a novel Graph Convolutional Neural Networks approach (GCNs-Net) on two EEG MI benchmarks, it has achieved dominant performances, 98.72\% accuracy on the PhysioNet Dataset, and 96.24\% on the High Gamma Dataset, for EEG MI decoding, which were far ahead than the results produced by the CNNs-based approaches. Moreover, more and more neuroscience research has suggested that the topological information promotes the analysis of brain network dynamics~\cite{bassett2017network, lv2014holistic, lv2017n}. Although the Euclidean distance was one of the similarity measurements, it might be superior to decode EEG signals from the non-Euclidean perspective by taking into consideration of the functional topological relationship of electrodes (\eg, the correlations and degree properties between electrodes) to enhance the decoding performance of EEG tasks.
	
	\begin{figure*}[h]
		\centering
		\includegraphics[width=\linewidth]{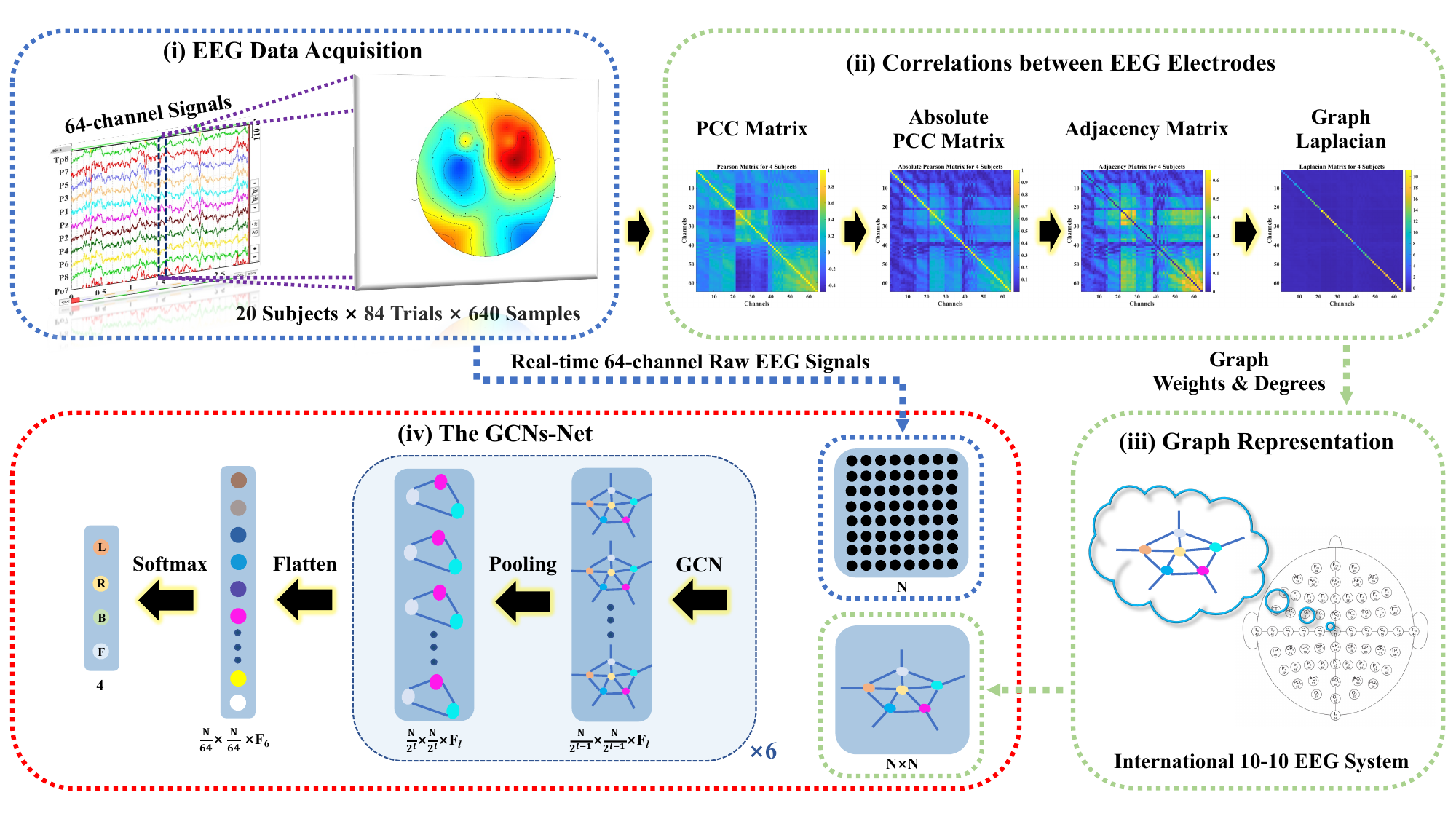}
		\caption{The system framework is composed of (i) acquisition of 64-channel raw EEG signals, (ii) correlation analysis for graph weights and degrees by presenting the PCC matrix, absolute PCC matrix, adjacency matrix, and graph Laplacian, (iii) graph representation of the International 10-10 EEG System, and (iv) a novel deep learning framework of the GCNs (the GCNs-Net).}
		\label{Overview}
	\end{figure*}
	
	Considering the topological relationship of EEG electrodes, the graph in the non-Euclidean space was put forward~\cite{zhang2019gcb, wang2019phase, song2018eeg, wang2018eeg, hou2022deep, jia2020attention}. Researchers have explored the CNNs with the graph theory, intending to apply the convolutional operation on graphs. Two strategies were presented to define convolutions,~\ie, either from the spatial domain or the spectral domain. At first, the spatial GCNs were proposed~\cite{hamilton2017inductive, monti2017geometric, niepert2016learning, gao2018large}. However, references~\cite{defferrard2016convolutional} and~\cite{bruna2013spectral} indicated that they faced the challenge of matching the local neighborhoods. On the other hand, the spectral method provided a well-defined localized operator on graphs~\cite{shuman2013emerging}. Thus, an innovative approach of the GCNs was proposed by implementing the CNNs based on the spectral graph theory. In specific, Bruna~\etal\cite{bruna2013spectral} first performed the GCNs from the spectral perspective. A few works~\cite{zhang2019gcb, wang2019phase, song2018eeg, wang2018eeg} have applied the above model to decode EEG tasks in the field of EEG-based emotion recognition. In detail, reference~\cite{zhang2019gcb} combined the GCNs with a broad learning system and put forward the Graph Convolutional Broad Network, which achieved 94.24\% accuracy on the SJTU Emotion EEG Dataset~\cite{7104132}. In~\cite{wang2019phase}, Wang~\etal presented a phase-locking value-based GCNs. The dynamical GCNs were proposed by Song~\cite{song2018eeg}, which can dynamically learn the topological relationship of EEG electrodes during training. Reference~\cite{wang2018eeg} improved the above method via a broad learning system. Moreover, in~\cite{hou2022deep}, Hou~\etal introduced a two-stage method in which it first extracted the spatial-temporal EEG features via an attention-based bidirectional long short-term memory model, and then employed the graph representation learning to classify deep features. Reference~\cite{jia2020attention} proposed an end-to-end approach, a graph residual network, to classify EEG signals. The highly competitive performance of these approaches~\cite{zhang2019gcb, wang2019phase, song2018eeg, wang2018eeg, hou2022deep, jia2020attention} indicated the superiority of applying the graph learning methods.
	
	In this work, a novel structure of the GCNs is introduced to decode EEG MI signals. First of all, based on the absolute Pearson’s matrix (PCC) of overall signals, the graph Laplacian is built up to represent the topological relationship of EEG electrodes. Besides, the GCNs-Net built on the graph convolutional layers learns the generalized features. The followed pooling layers reduce dimensionality. And the fully-connected (FC) softmax layer derives the final prediction. Furthermore, the Chebyshev polynomial is applied to approximate the graph convolutional filters, which significantly promotes computational efficiency. Last but not least, the GCNs-Net decodes time-resolved EEG MI signals, which paves the road towards effective and efficient BCI applications. The main contributions of this work are summarized as follows.
	\begin{enumerate}
		\item A novel structure of the GCNs is introduced to detect four-class MI intentions while cooperating with the functional topological relationship of EEG electrodes.
		\item The individual and group-wise performance of the GCNs-Net on two benchmark datasets of EEG MI outperforms the existing studies, which validates that the method can decode relevant features for brain motor imagination.
		\item The introduced GCNs-Net framework can be easily transferred and implemented for other MI-related tasks, and potentially for other EEG BCI tasks.
	\end{enumerate}
	
	\section{Materials and Methods}\label{Materials and Methods}
	\subsection{Overview}\label{Overview_sec}
	The framework of this work is shown in~\reffig{Overview}.
	\begin{enumerate}[label={(\roman*)}]
		\item 64-channel raw EEG signals are acquired as one of the inputs of the GCNs-Net.
		\item The PCC matrix, absolute PCC matrix, adjacency matrix, and graph Laplacian are introduced to represent the correlations between electrodes.
		\item The graph representation, another input of the GCNs-Net, is represented by the graph Laplacian.
		\item The GCNs-Net is applied to decode EEG MI signals, where $\rm N$ denotes the number of electrodes and $l$ denotes the $l^{th}$ graph pooling layer.
	\end{enumerate}
	
	\subsection{Dataset Description}\label{Dataset Description}
	In this work, two benchmark datasets are used to evaluate the effectiveness and robustness of our presented method. 
	
	\textbf{The PhysioNet Dataset}: The EEG Motor Movement/Imagery Dataset consists of over 1,500 EEG records from 109 subjects~\cite{goldberger2000physiobank}. There are 64 electrodes based on the international 10-10 system. Each subject performs 84 trials (3 runs $\times$ 7 trials $\times$ 4 tasks). 160 Hz sampling rate and 4-second signals,~\ie, 640 time points per trail, are utilized considering the duration of the experiments~\cite{hou2019novel}. Four MI tasks are termed as L (image left fist), R (image right fist), B (image both fists), and F (image both feet), respectively. 
	
	\textbf{The High Gamma Dataset}: Collected from 14 subjects, the Public High Gamma Dataset performs four EEG tasks,~\ie, left-hand movement, right-hand movement, both feet movement, and rest~\cite{tangermann2012review}. The data of 44 electrodes and 0-125 Hz frequency are applied in our experiments, and the dataset is resampled to 250 Hz~\cite{schirrmeister2017deep}. Each subject performs approximately 880 trials for training and 160 trials for testing.
	
	Previous studies apply segments (windows) of time points as samples~\cite{hou2022deep, jiao2018deep, chai2017improving}. However, neuroscience research indicates that the brain is one of the most complicated systems, and its state at every moment is changing. Time-resolved signals can represent the condition of the brain at the instant moment, which reflects the network patterns of brain dynamics. Therefore, in this work, every time point is recognized as a sample to map the brain state to the corresponding MI task. Compared with applying the time-window signals as samples, the method is time-resolved, which is superior to pave the road towards developing real-time and efficient EEG MI applications~\cite{jia2020attention}.
	
	In this paper, multiple subjects (S$_1$$\sim$S$_{20}$, S$_1$$\sim$S$_{50}$, S$_1$$\sim$S$_{100}$) from the PhysioNet Dataset and 14 subjects from the High Gamma Dataset are picked up to train and evaluate the GCNs-Net~\cite{hou2019novel}. Over all the experiments, 90\% of the dataset is chosen as the training set, and the remaining 10\% serves as the testing set, following the default subject-dependent scenario~\cite{zhang2019making, hou2019novel} for fair performance comparison. And we repeat the above procedure 10 times with different random seeds to validate the stability and reliability of the GCNs-Net and avoid obtaining bias results with pseudo-random. Then, we use the averaged accuracy as the final performance of the model. For the group-level analysis, we carefully divide the dataset for each subject to prevent data imbalance and model bias.
	
	\subsection{Graph Preliminary}\label{Graph Preliminary}
	\subsubsection{Graph Representation}\label{Gragh Representation} 
	An undirected and weighted graph is represented by $\boldsymbol{\rm G}=\{{\boldsymbol{\rm V}, \boldsymbol{\rm E}, \boldsymbol{\rm A}\}}$, in which $\boldsymbol{\rm V}$ denotes a set of nodes with the number $\vert\boldsymbol{\rm V}\vert = \emph{N}$, $\boldsymbol{\rm E}$ denotes a set of edges connecting nodes, and $\boldsymbol{\rm A} \in \mathbb{R}\textsuperscript{\emph{N}$\times$\emph{N}}$ is a weighted adjacency matrix representing correlations between two nodes.
	
\begin{figure*}[ht]
		\centering
		\begin{minipage}[t]{.24\linewidth}
			\includegraphics[width=1.65in]{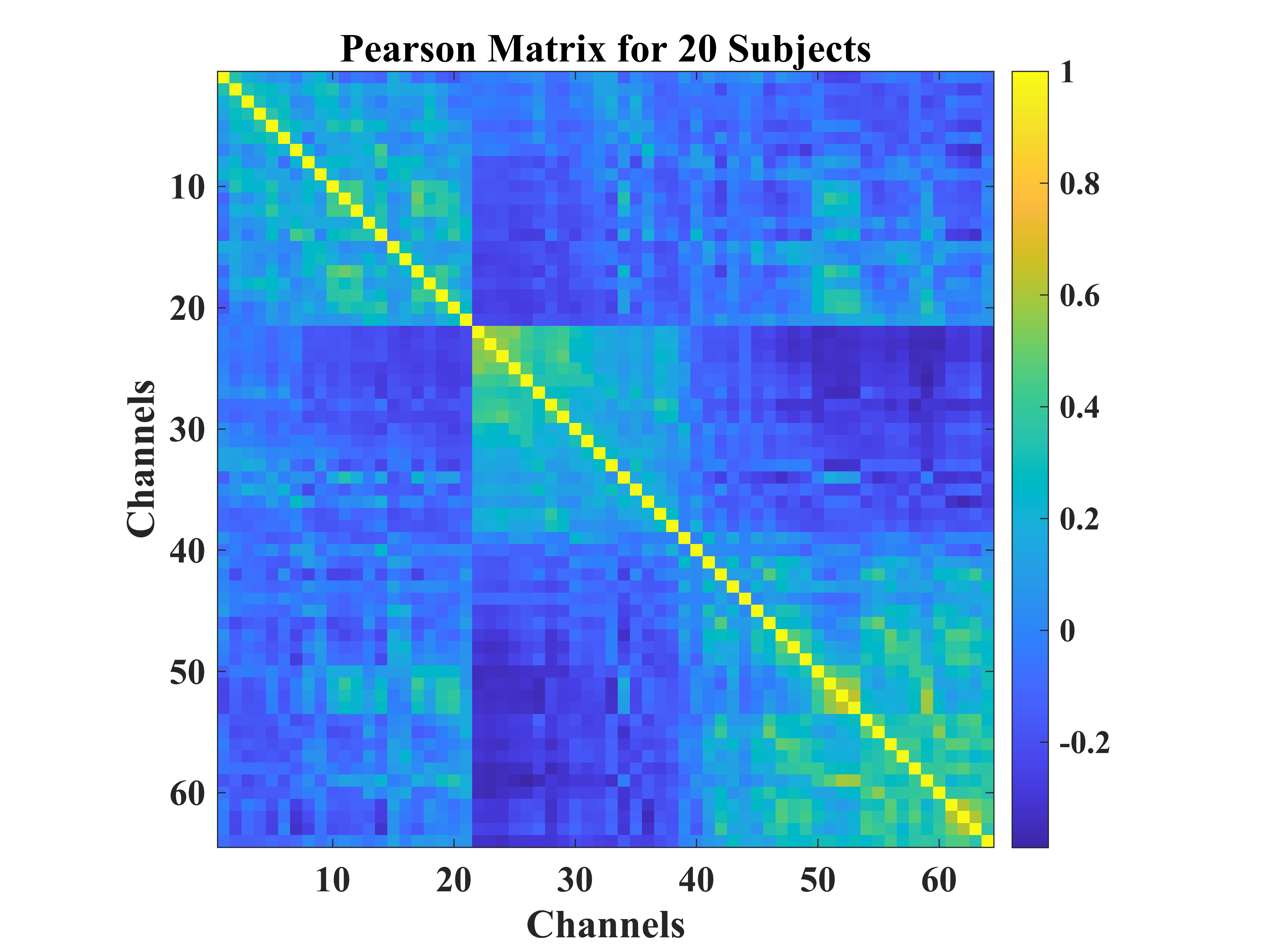}
			\subcaption{}
			\label{Pearson_matrix_for_20_Subjects}
		\end{minipage}
		\begin{minipage}[t]{.24\linewidth}
			\includegraphics[width=1.65in]{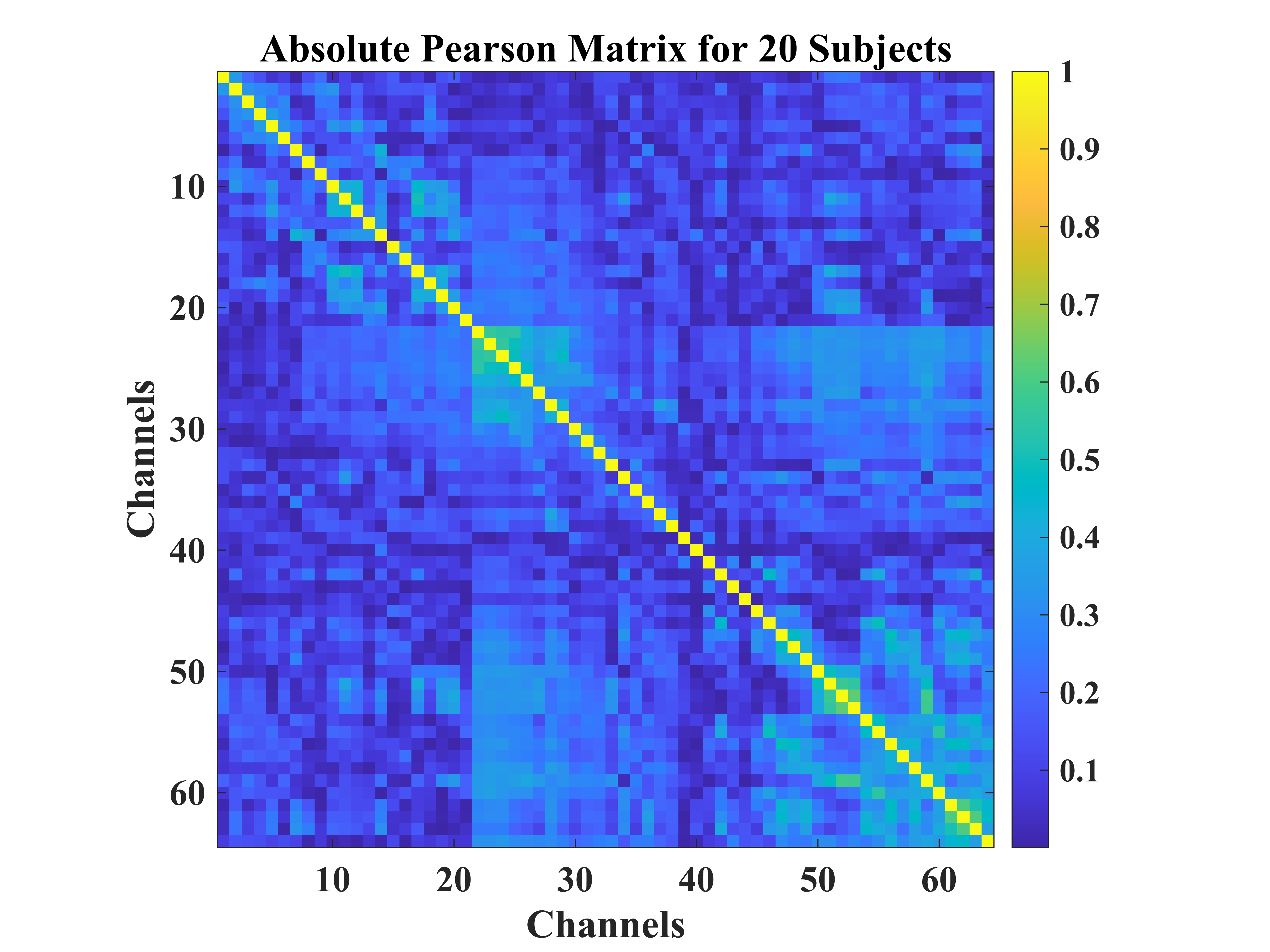}
			\subcaption{}
			\label{Absolute_Pearson_matrix_for_20_Subjects}
		\end{minipage}
		\begin{minipage}[t]{.24\linewidth}
			\includegraphics[width=1.65in]{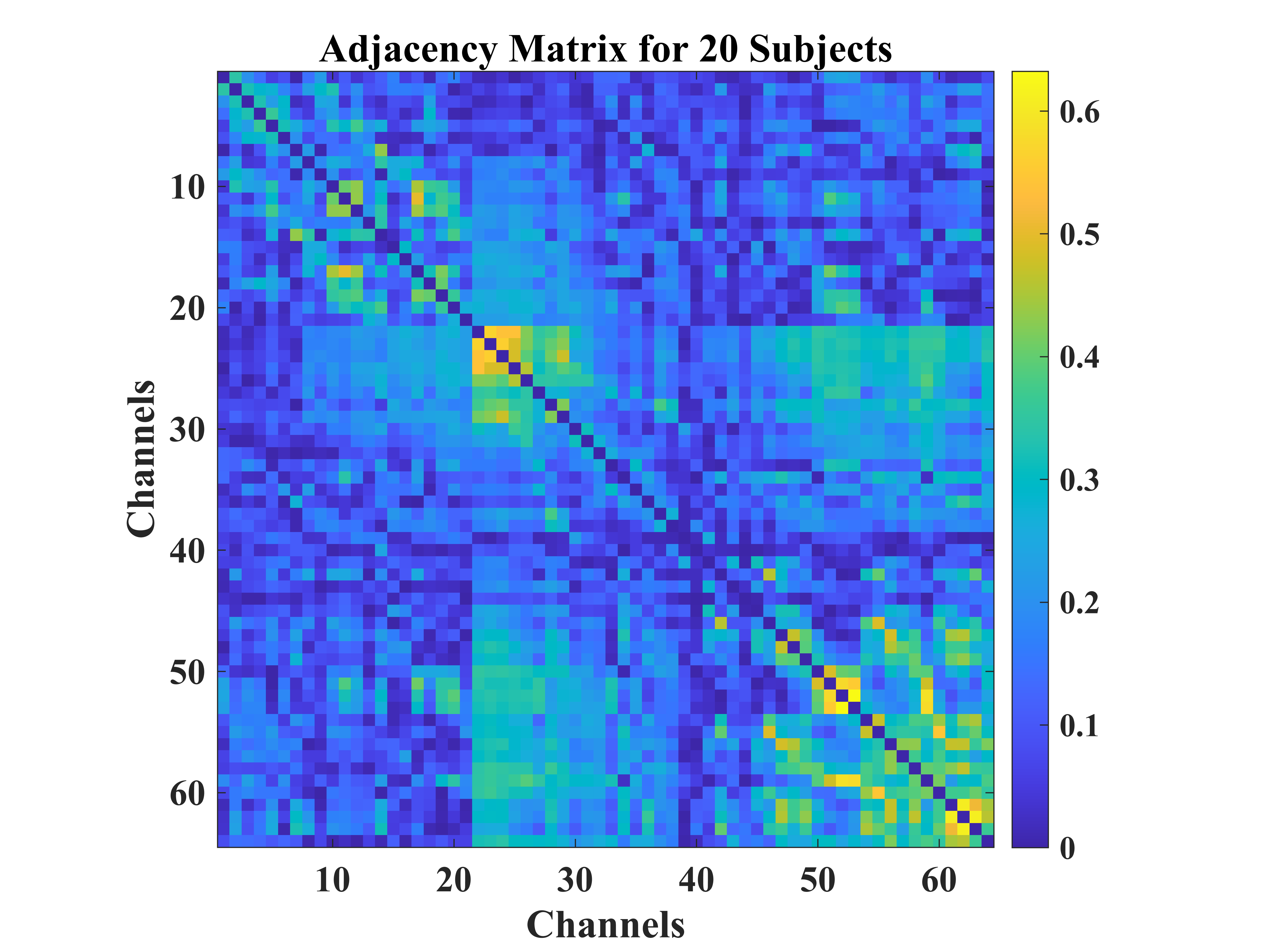}
			\subcaption{}
			\label{Adjacency_Matrix_for_20_Subjects}
		\end{minipage}
		\begin{minipage}[t]{.24\linewidth}
			\includegraphics[width=1.65in]{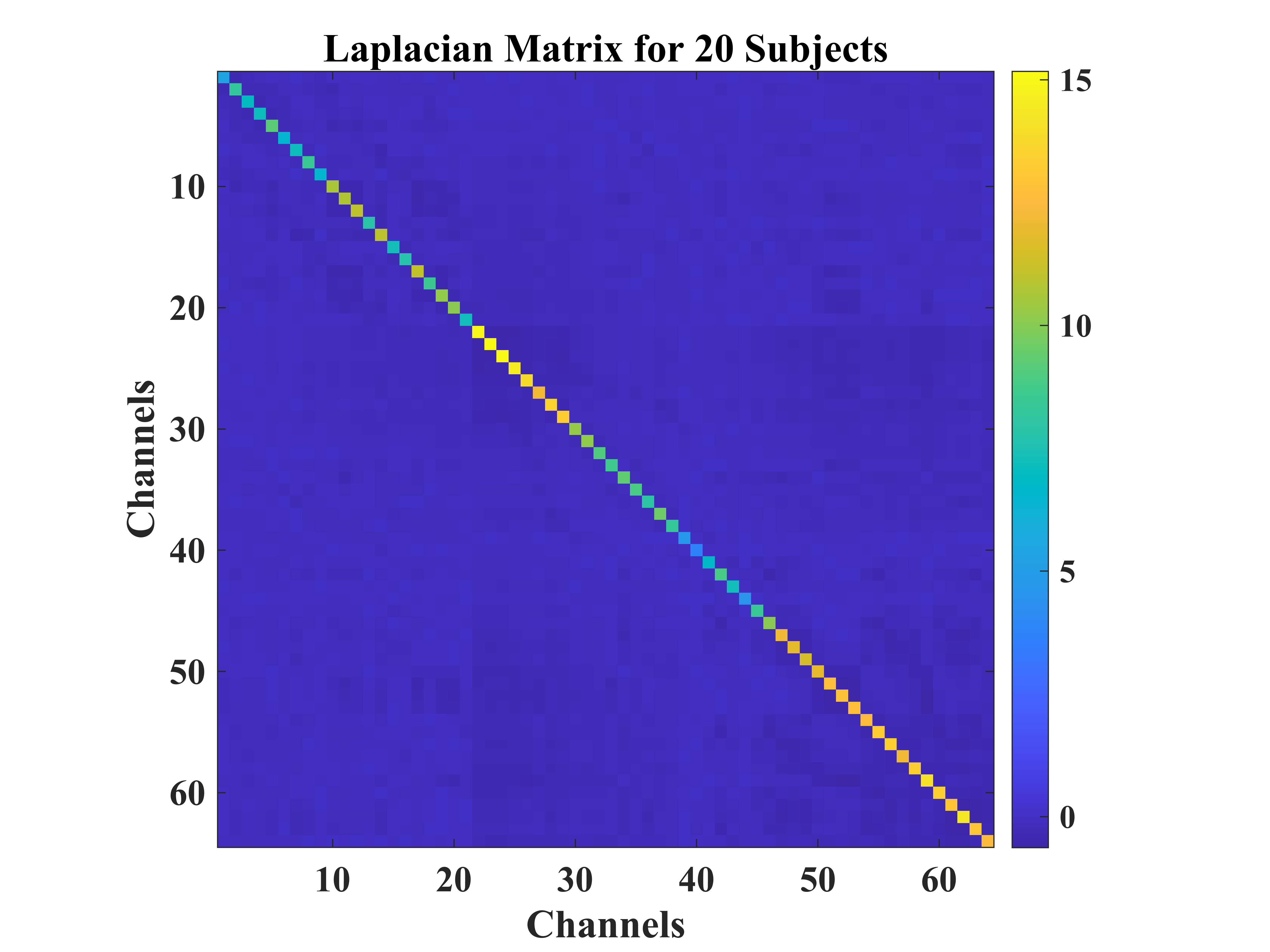}
			\subcaption{}
			\label{Laplacian_Matrix_for_20_Subjects}
		\end{minipage}
		\begin{minipage}[t]{.24\linewidth}
			\includegraphics[width=1.65in]{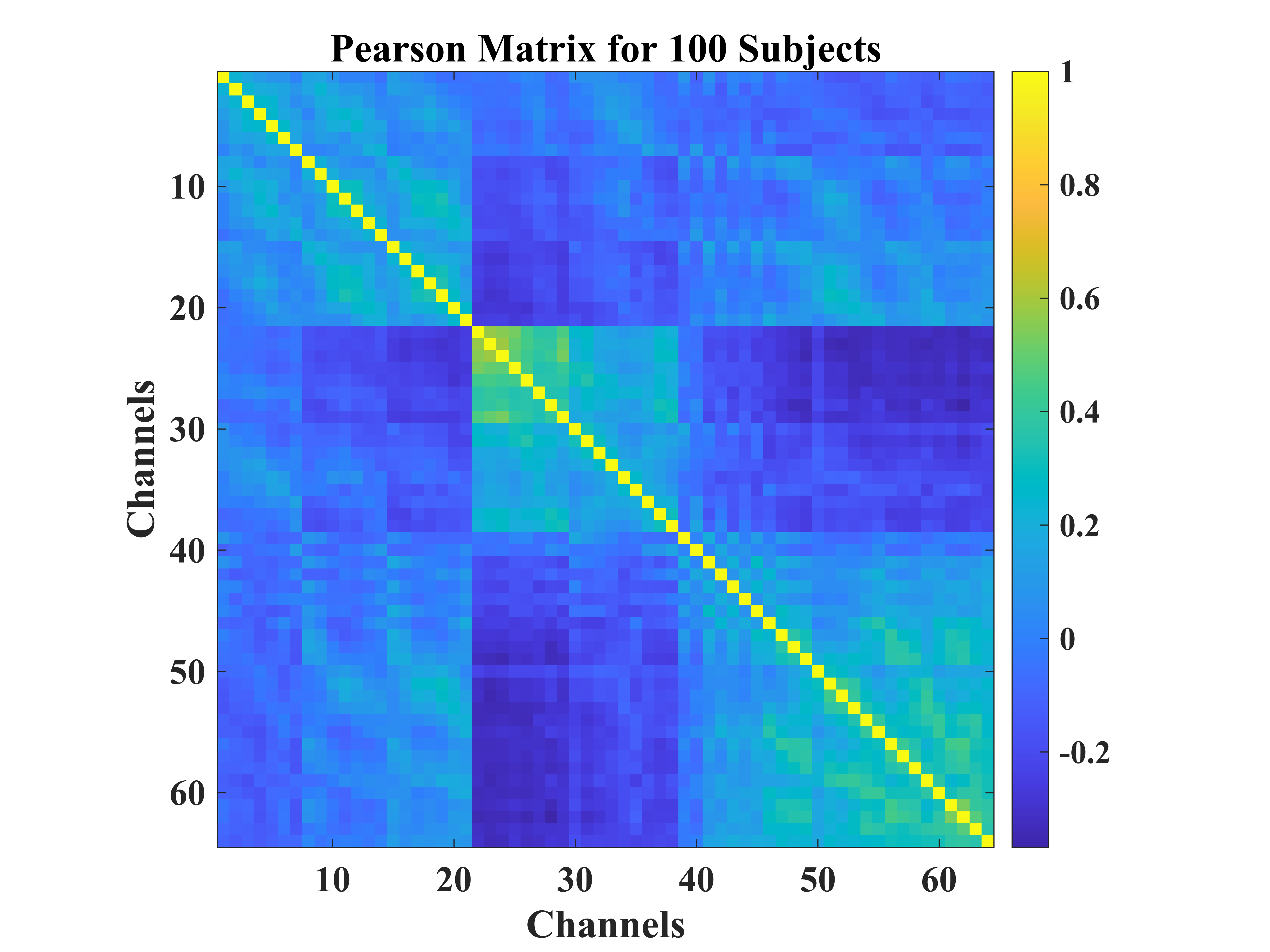}
			\subcaption{}
			\label{Pearson_matrix_for_100_Subjects}
		\end{minipage}
		\begin{minipage}[t]{.24\linewidth}
			\includegraphics[width=1.65in]{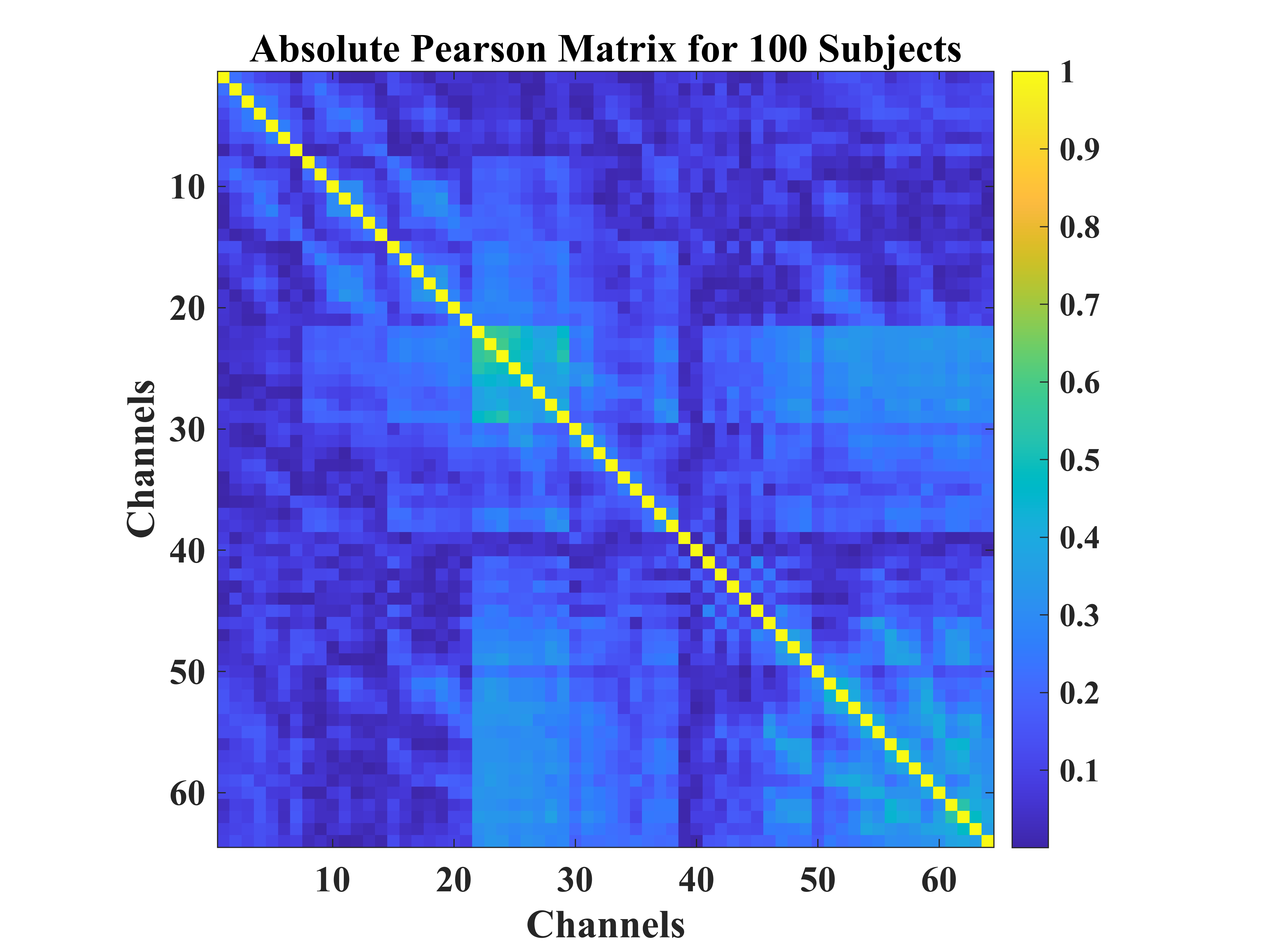}
			\subcaption{}
			\label{Absolute_Pearson_matrix_for_100_Subjects}
		\end{minipage}
		\begin{minipage}[t]{.24\linewidth}
			\includegraphics[width=1.65in]{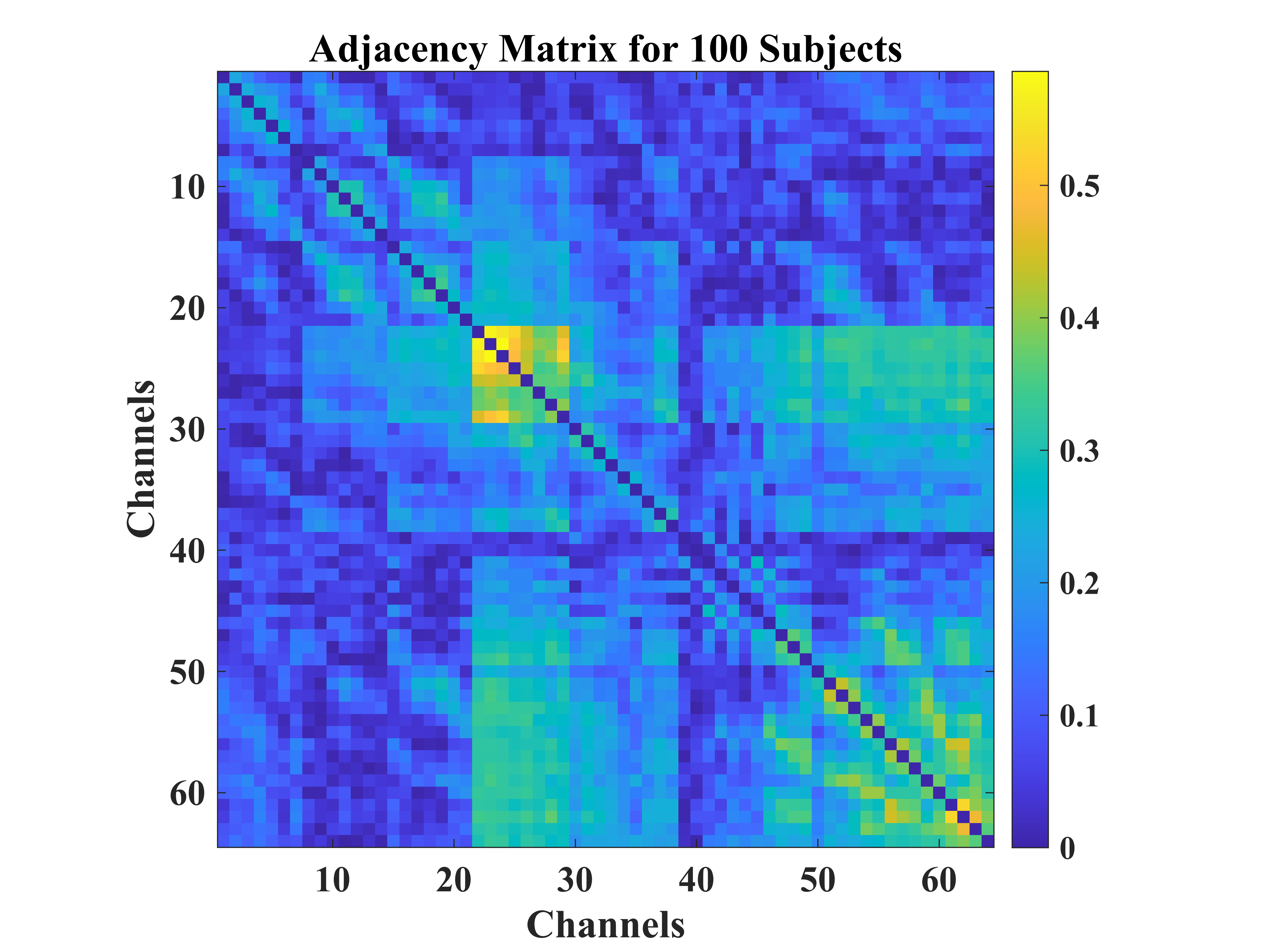}
			\subcaption{}
			\label{Adjacency_Matrix_for_100_Subjects}
		\end{minipage}
		\begin{minipage}[t]{.24\linewidth}
			\includegraphics[width=1.65in]{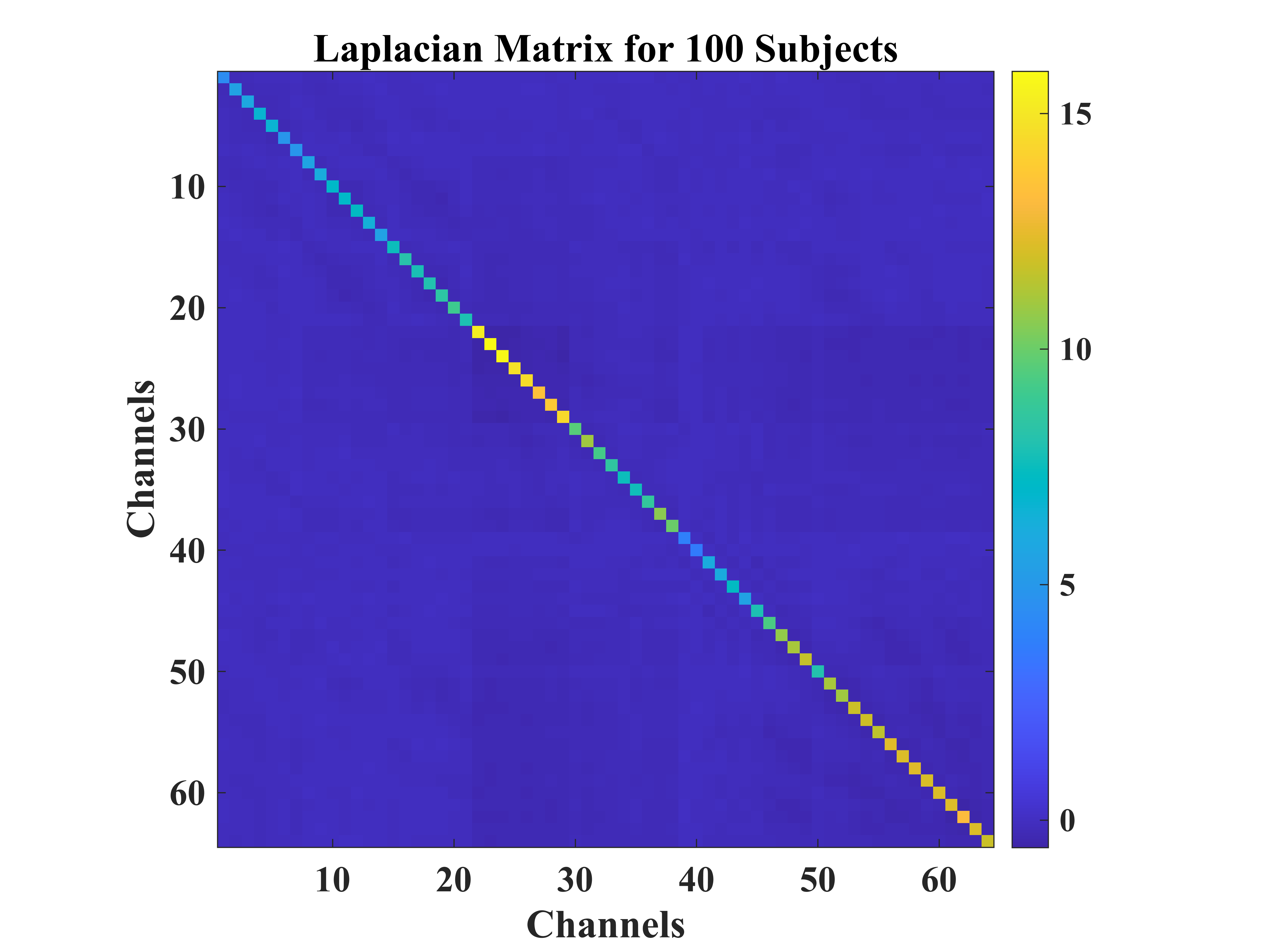}
			\subcaption{}
			\label{Laplacian_Matrix_for_100_Subjects}
		\end{minipage}
		\caption{The PCC matrix, absolute PCC matrix, adjacency matrix, and graph Laplacian for 20 and 100 subjects, respectively, from the PhysioNet Dataset. (a) The PCC matrix for 20 subjects. (b) The absolute PCC matrix for 20 subjects. (c) The adjacency matrix for 20 subjects. (d) The graph Laplacian for 20 subjects. (e) The PCC matrix for 100 subjects. (f) The absolute PCC matrix for 100 subjects. (g) The adjacency matrix for 100 subjects. (h) The graph Laplacian for 100 subjects.}
		\label{PCC}
	\end{figure*}

	To present the degree matrix of a graph, the scale of the graph weights is analysed regardless of the polar relevance,~\ie, whether the correlations are positive or negative. So the absolute Pearson Correlation Coefficient (PCC) matrix $\vert\boldsymbol{\rm P}\vert \in [0, 1]$ is introduced, which is the absolute of the PCC matrix $\boldsymbol{\rm P}$, to map the linear correlations between signals. $\boldsymbol{\rm A}$ is represented as $\boldsymbol{\rm A}=\vert\boldsymbol{\rm P}\vert - \boldsymbol{\rm I}$, where $\boldsymbol{\rm I}$ is an identity matrix. The PCC matrix, absolute PCC matrix, adjacency matrix, and graph Laplacian for 20 and 100 subjects are given in~\reffig{PCC}, respectively.
	
	\reffig{Pearson_matrix_for_20_Subjects} and~\reffig{Pearson_matrix_for_100_Subjects} demonstrate that there are no apparent differences on the correlations of various sizes of datasets. The reason for this phenomenon is that all the signals of participants are utilized to compute the PCC matrix. Furthermore, the degree matrix $\boldsymbol{\rm D}$ is obtained, which is a diagonal matrix, and the $\it{i}$-th diagonal element can be computed by: $D_{ii}=\sum_{j=1}^N\boldsymbol{\rm A}_{ij}$. Finally, the combinatorial Laplacian $\boldsymbol{\rm L} \in \mathbb{R}\textsuperscript{\emph{N}$\times$\emph{N}}$ is represented as $\boldsymbol{\rm L}=\boldsymbol{\rm D}-\boldsymbol{\rm A}$. The graph Laplacians for 20 and 100 subjects are presented in~\reffig{Laplacian_Matrix_for_20_Subjects} and~\reffig{Laplacian_Matrix_for_100_Subjects}, respectively. And the normalized graph Laplacian $\boldsymbol{\rm L}=\boldsymbol{\rm I}_{N}-\boldsymbol{\rm D}^{-1 / 2}\boldsymbol{\rm A}\boldsymbol{\rm D}^{-1 / 2}$ represents the correlations between nodes.
	
	\subsubsection{Spectral Graph Filtering}\label{Spectral Graph Filtering}
	As $\boldsymbol{\rm L}$ is a real symmetric positive semidefinite matrix, its set of eigenvectors,~\ie, the graph Fourier modes, $\left\{\boldsymbol{\rm u_{l}}\right\}_{l=0}^{N-1}\in\mathbb{R}^{N}$, are complete and orthonormal. And the associated eigenvalues, as known as the graph Fourier frequencies, $\left\{\lambda_{l}\right\}_{l=0}^{N-1}\in\mathbb{R}$, are ordered and real nonnegative. The Fourier basis $\boldsymbol{\rm U}=\left[\boldsymbol{\rm u_{0}}, \ldots, \boldsymbol{\rm u_{N-1}}\right] \in \mathbb{R}^{N \times N}$ decomposes the graph Laplacian,~\ie, $\boldsymbol{\rm L}=\boldsymbol{\rm U}\Lambda \boldsymbol{\rm U}^{T}$, where $\boldsymbol{\rm \Lambda}=\mathrm{diag}\left(\left[\lambda_{0}, \ldots, \lambda_{N-1}\right]\right) \in \mathbb{R}^{N \times N}$. The signal $\boldsymbol{\rm x} \in \mathbb{R}^{N}$ transformed by the graph Fourier is represented as $\hat{\boldsymbol{\rm x}}=\boldsymbol{\rm U}^{T}\boldsymbol{\rm x}\in \mathbb{R}^{N}$, and its inverse graph Fourier transform is $\boldsymbol{\rm x}=\boldsymbol{\rm U}\hat{\boldsymbol{\rm x}}$~\cite{shuman2013emerging}. It has projected the input signals to an orthonormal space where the bases are formed by the eigenvectors of the normalized graph Laplacian~\cite{wu2019comprehensive}.
	
	The convolution on graph $\boldsymbol{\rm G}$ is defined below:
	\begin{equation}
		\boldsymbol{\rm x} *_{\boldsymbol{\rm G}} \boldsymbol{\rm g}=\boldsymbol{\rm U}\left(\left(\boldsymbol{\rm U}^{T} \boldsymbol{\rm x}\right) \odot\left(\boldsymbol{\rm U}^{T} \boldsymbol{\rm g}\right)\right),
	\end{equation}
	in which $\odot$ denotes the element-wise Hadamard product and $\boldsymbol{\rm g}\in\mathbb{R}^{N}$ is a convolutional filter. Here, $\boldsymbol{\rm g}$ is non-parametric and it is denoted as $\boldsymbol{\rm g}_{\theta}(\boldsymbol{\rm \Lambda})=\mathrm{diag}(\boldsymbol{\rm \theta})$, where $\boldsymbol{\rm \theta}\in\mathbb{R}^{N}$ is the vector of the Fourier coefficients. The convolutional operation implemented in the GCNs is in the following.
	\begin{equation}
		\boldsymbol{\rm x} *_{\boldsymbol{\rm G}} \boldsymbol{\rm g}_{\theta}=\boldsymbol{\rm g}_{\theta}\left(\boldsymbol{\rm U} \boldsymbol{\rm \Lambda} \boldsymbol{\rm U}^{T}\right) \boldsymbol{\rm x}=\boldsymbol{\rm U} \boldsymbol{\rm g}_{\theta}(\boldsymbol{\rm \Lambda}) \boldsymbol{\rm U}^{T} \boldsymbol{\rm x}.
	\end{equation}
	
	\begin{table*}[!t]
		\centering
		\caption{Implementation details of the proposed GCNs-Net on the PhysioNet Dataset}
		\label{Implementation details for proposed GCNs-Net}
		\resizebox{\linewidth}{!}{
			\begin{tabular}{cccccccccc}
				\toprule
				Layer & Type & Maps & Size & Edges & \tabincell{c}{Polynomial \\ Order} & \tabincell{c}{Pooling \\ Size} & Activation & Weights & Bias \\ \midrule
				Softmax & Fully-connected & $-$ & $\rm O$ & $-$ & $-$ & $-$ & Softmax & $\frac{\rm N}{64}\times\frac{\rm N}{64}\times{\rm F}_{6}\times{\rm O}$ & $\rm O$ \\ 
				Flatten & Flatten & $-$ & $\frac{\rm N}{64}\times\frac{\rm N}{64}\times{\rm F}_{6}$ & $-$ & $-$ & $-$ & $-$ & $-$ & $-$ \\ 
				P6 & Max-pooling & ${\rm F}_{6}$ & $\frac{\rm N}{32}$ & $\sum\nolimits_{i=1}^{\frac{\rm N}{32}-1}i$ & $-$ & 2 & $-$ & $-$ & $-$ \\ 
				C6 & Convolution & ${\rm F}_{6}$ & $\frac{\rm N}{32}$ & $\sum\nolimits_{i=1}^{\frac{\rm N}{32}-1}i$ & $\rm K$ & $-$ & Softplus & 
				${\rm F}_{5}\times{\rm F}_{6}\times{\rm K}$ & $\frac{\rm N}{32}\times{\rm F}_{6}$ \\ 
				P5 & Max-pooling & ${\rm F}_{5}$ & $\frac{\rm N}{16}$ & $\sum\nolimits_{i=1}^{\frac{\rm N}{16}-1}i$ & $-$ & 2 & $-$ & $-$ & $-$ \\ 
				C5 & Convolution & ${\rm F}_{5}$ & $\frac{\rm N}{16}$ & $\sum\nolimits_{i=1}^{\frac{\rm N}{16}-1}i$ & $\rm K$ & $-$ & Softplus & ${\rm F}_{4}\times{\rm F}_{5}\times{\rm K}$ & $\frac{\rm N}{16}\times{\rm F}_{5}$ \\ 
				P4 & Max-pooling & ${\rm F}_{4}$ & $\frac{\rm N}{8}$ & $\sum\nolimits_{i=1}^{\frac{\rm N}{8}-1}i$ & $-$ & 2 & $-$ & $-$ & $-$ \\ 
				C4 & Convolution & ${\rm F}_{4}$ & $\frac{\rm N}{8}$ & $\sum\nolimits_{i=1}^{\frac{\rm N}{8}-1}i$ & $\rm K$ & $-$ & Softplus & ${\rm F}_{3}\times{\rm F}_{4}\times{\rm K}$ & $\frac{\rm N}{8}\times{\rm F}_{4}$ \\ 
				P3 & Max-pooling & ${\rm F}_{3}$ & $\frac{\rm N}{4}$ & $\sum\nolimits_{i=1}^{\frac{\rm N}{4}-1}i$ & $-$ & 2 & $-$ & $-$ & $-$ \\ 
				C3 & Convolution & ${\rm F}_{3}$ & $\frac{\rm N}{4}$ & $\sum\nolimits_{i=1}^{\frac{\rm N}{4}-1}i$ & $\rm K$ & $-$ & Softplus & ${\rm F}_{2}\times{\rm F}_{3}\times{\rm K}$ & $\frac{\rm N}{4}\times{\rm F}_{3}$ \\ 
				P2 & Max-pooling & ${\rm F}_{2}$ & $\frac{\rm N}{2}$ & $\sum\nolimits_{i=1}^{\frac{\rm N}{2}-1}i$ & $-$ & 2 & $-$ & $-$ & $-$ \\ 
				C2 & Convolution & ${\rm F}_{2}$ & $\frac{\rm N}{2}$ & $\sum\nolimits_{i=1}^{\frac{\rm N}{2}-1}i$ & $\rm K$ & $-$ & Softplus & ${\rm F}_{1}\times{\rm F}_{2}\times{\rm K}$ & $\frac{\rm N}{2}\times{\rm F}_{2}$ \\ 
				P1 & Max-pooling & ${\rm F}_{1}$ & $\rm N$ & $\sum\nolimits_{i=1}^{N-1}i$ & $-$ & 2 & $-$ & $-$ & $-$ \\ 
				C1 & Convolution & ${\rm F}_{1}$ & $\rm N$ & $\sum\nolimits_{i=1}^{N-1}i$ & $\rm K$ & $-$ & Softplus & $1\times{\rm F}_{1}\times{\rm K}$ & ${\rm N}\times{\rm F}_{1}$ \\ 
				Input & Input & 1 & $\rm N$ & $\sum\nolimits_{i=1}^{N-1}i$ & $-$ & $-$ & $-$ & $-$ & $-$ \\ 
				\bottomrule
			\end{tabular}
		}
	\end{table*}
	
	The difference of the spectral graph convolution lies in the choice of the filter $\boldsymbol{\rm g}_{\theta}$. Since the non-parametric filter is not localized in space, and its computational complexity is too high, we use the polynomial approximation to address the problem. The Chebyshev polynomials are popularly utilized to approximate filters~\cite{hammond2011wavelets}. $\boldsymbol{\rm g}_{\theta}$ is parametrized as a truncated expansion as follows:
	\begin{equation}
		\boldsymbol{\rm g}_{\theta}(\boldsymbol{\rm \Lambda})=\sum_{k=0}^{K-1} \theta_{k} \boldsymbol{\rm T}_{k}(\tilde{\boldsymbol{\rm \Lambda}}),
	\end{equation}
	in which parameters $\boldsymbol{\rm \theta} \in \mathbb{R}^{K}$ is a set of Chebyshev coefficients, $\boldsymbol{\rm T}_{k}(\tilde{\boldsymbol{\rm \Lambda}}) \in \mathbb{R}^{K}$ is the $k^{th}$ order Chebyshev polynomial evaluated at $\tilde{\boldsymbol{\rm \Lambda}}=2 \boldsymbol{\rm \Lambda} / \Lambda_{max}-\boldsymbol{\rm I}_{N}$, and $\boldsymbol{\rm I}_{N}$ is a diagonal matrix of the scaled eigenvalues.
	
	Then, the signal $\boldsymbol{\rm x}$ is convolutioned by the defined filter $\boldsymbol{\rm g}_{\theta}$ as follows:
	\begin{equation}
		\boldsymbol{\rm x} *_{\boldsymbol{\rm G}} \boldsymbol{\rm g}_{\theta}=\boldsymbol{\rm U} \sum_{k=0}^{K-1} \theta_{k} \boldsymbol{\rm T}_{k}(\tilde{\boldsymbol{\rm \Lambda}}) \boldsymbol{\rm U}^{T} \boldsymbol{\rm x}=\sum_{k=0}^{K-1} \theta_{k} \boldsymbol{\rm T}_{k}(\tilde{\boldsymbol{\rm L}}) \boldsymbol{\rm x}.
	\end{equation}
	$\boldsymbol{\rm T}_{k}(\tilde{\boldsymbol{\rm L}})$ is the Chebyshev polynomial of order $k$ evaluated at the scaled Laplacian $\tilde{\boldsymbol{\rm L}}=2 \boldsymbol{\rm L} / \lambda_{max}-\boldsymbol{\rm I}_{N}$. Let $\bar{\boldsymbol{\rm x}}_{k}=\boldsymbol{\rm T}_{k}(\tilde{\boldsymbol{\rm L}}) \boldsymbol{\rm x} \in \mathbb{R}^{N}$, a recursive relation is utilized to compute $\bar{\boldsymbol{\rm x}}_{k}$,~\ie, $\bar{\boldsymbol{\rm x}}_{k}=2 \tilde{\boldsymbol{\rm L}} \bar{\boldsymbol{\rm x}}_{k-1}-\bar{\boldsymbol{\rm x}}_{k-2}$ with $\bar{\boldsymbol{\rm x}}_{0}=\boldsymbol{\rm x}$, and $\bar{\boldsymbol{\rm x}}_{1}=\tilde{\boldsymbol{\rm L}} \boldsymbol{\rm x}$. Another reason why the Chebyshev polynomial is applied to approximate convolutional filters is that it implicitly avoids computations for the graph Fourier basis, thus reducing the computational complexity from $O(N^2)$ to $O(KN)$. 
	\begin{table*}[!ht]
		\centering
		\caption{Performance comparisons of the GCNs-Net}
		\resizebox{\linewidth}{!}{
			\begin{tabular}{cccccccccc}
				\toprule
				Model & \tabincell{c}{Num. of \\ Conv Layers} & \tabincell{c}{Num. of \\ Pooling Layers} & \tabincell{c}{Num. of \\ Filters} & Model Framework & \tabincell{c}{Accuracy w.r.t. \\ $1^{st} order$} & \tabincell{c}{Accuracy w.r.t. \\ $2^{nd} order$} & \tabincell{c}{Accuracy w.r.t. \\ $3^{rd} order$} & \tabincell{c}{Accuracy w.r.t. \\ $4^{th} order$} & \tabincell{c}{Accuracy w.r.t. \\ $5^{th} order$}\\ \midrule
				1 & 1 & 1 & 16 & C-P-S & 55.63\% & 55.30\% & 56.70\% & 56.60\% & 56.04\% \\ 
				2 & 2 & 1 & 16, 32 & C-C-P-S & 57.94\% & 61.90\% & 63.17\% & 63.16\% & 63.37\% \\ 
				3 & 2 & 2 & 16, 32 & (C-P)$\times$2-S & 60.04\% & 62.32\% & 62.55\% & 62.96\% & 62.08\% \\ 
				4 & 3 & 1 & 16, 32, 64 & C-C-C-P-S & 58.07\% & 69.18\% & 69.86\% &71.17\% & 70.98\% \\ 
				5 & 3 & 2 & 16, 32, 64 & C-(C-P)$\times$2-S & 61.65\% & 69.73\% & 70.19\% & 71.06\% & 71.45\% \\ 
				6 & 3 & 3 & 16, 32, 64 & (C-P)$\times$3-S & 65.03\% & 70.50\% & 69.12\% & 70.36\% & 71.30\% \\ 
				7 & 4 & 2 & 16, 32, 64, 128 & C-C-(C-P)$\times$2-S & 63.27\% & 77.09\% & 77.04\% & 78.42\% & 78.15\% \\ 
				8 & 4 & 2 & 16, 32, 64, 128 & (C-C-P)$\times$2-S & 63.69\% & 77.59\% & 77.32\% & 79.28\% & 77.43\% \\ 
				9 & 4 & 3 & 16, 32, 64, 128 & C-(C-P)$\times$3-S & 67.50\% & 77.63\% & 77.67\% & 79.60\% & 78.36\% \\ 
				10 & 4 & 4 & 16, 32, 64, 128 & (C-P)$\times$4-S & 71.22\% & 77.61\% & 78.50\% & 78.22\% & 78.26\% \\ 
				11 & 5 & 3 & 16, 32, 64, 128, 256 & C-C-(C-P)$\times$3-S & 70.11\% & 83.06\% & 82.78\% & 84.29\% & 84.19\% \\ 
				12 & 5 & 3 & 16, 32, 64, 128, 256 & (C-C-P)$\times$2-C-P-S & 70.12\% & 83.05\% & 82.49\% & 84.26\% & 83.45\% \\ 
				13 & 5 & 4 & 16, 32, 64, 128, 256 & C-(C-P)$\times$4-S & 75.93\% & 84.17\% & 83.90\% & 84.74\% & 84.57\% \\ 
				14 & 5 & 5 & 16, 32, 64, 128, 256 & (C-P)$\times$5-S & 77.79\% & 84.39\% & 84.30\% & 83.90\% & 85.08\% \\ 
				15 & 6 & 3 & 16, 32, 64, 128, 256, 512 & C-C-C-(C-P)$\times$3-S & 70.73\% & 86.77\% & 86.52\% & 87.62\% & 87.30\% \\ 
				16 & 6 & 3 & 16, 32, 64, 128, 256, 512 & (C-C-P)$\times$3-S & 73.59\% & 87.09\% & 86.71\% & 87.83\% & 87.21\% \\ 
				17 & 6 & 4 & 16, 32, 64, 128, 256, 512 & C-C-(C-P)$\times$4-S & 77.69\% & 87.63\% & 87.18\% & 88.18\% & 87.94\% \\ 
				18 & 6 & 4 & 16, 32, 64, 128, 256, 512 & C-P-(C-C-P)$\times$2-C-P-S & 78.38\% & 87.65\% & 87.70\% & 87.80\% & 87.36\% \\ 
				19 & 6 & 5 & 16, 32, 64, 128, 256, 512 & C-(C-P)$\times$5-S & 81.89\% & 88.60\% & 88.08\% & 88.45\% & 88.97\% \\ 
				20 & 6 & 6 & 16, 32, 64, 128, 256, 512 & (C-P)$\times$6-S & 84.88\% & 88.85\% & 88.25\% & 88.37\% & 87.90\% \\ \bottomrule
			\end{tabular}
		}
		\label{Model framework in the experiments}
	\end{table*}
	\begin{figure*}[!ht]
		\centering
		\begin{minipage}[t]{.24\linewidth}
			\includegraphics[width=1.65in]{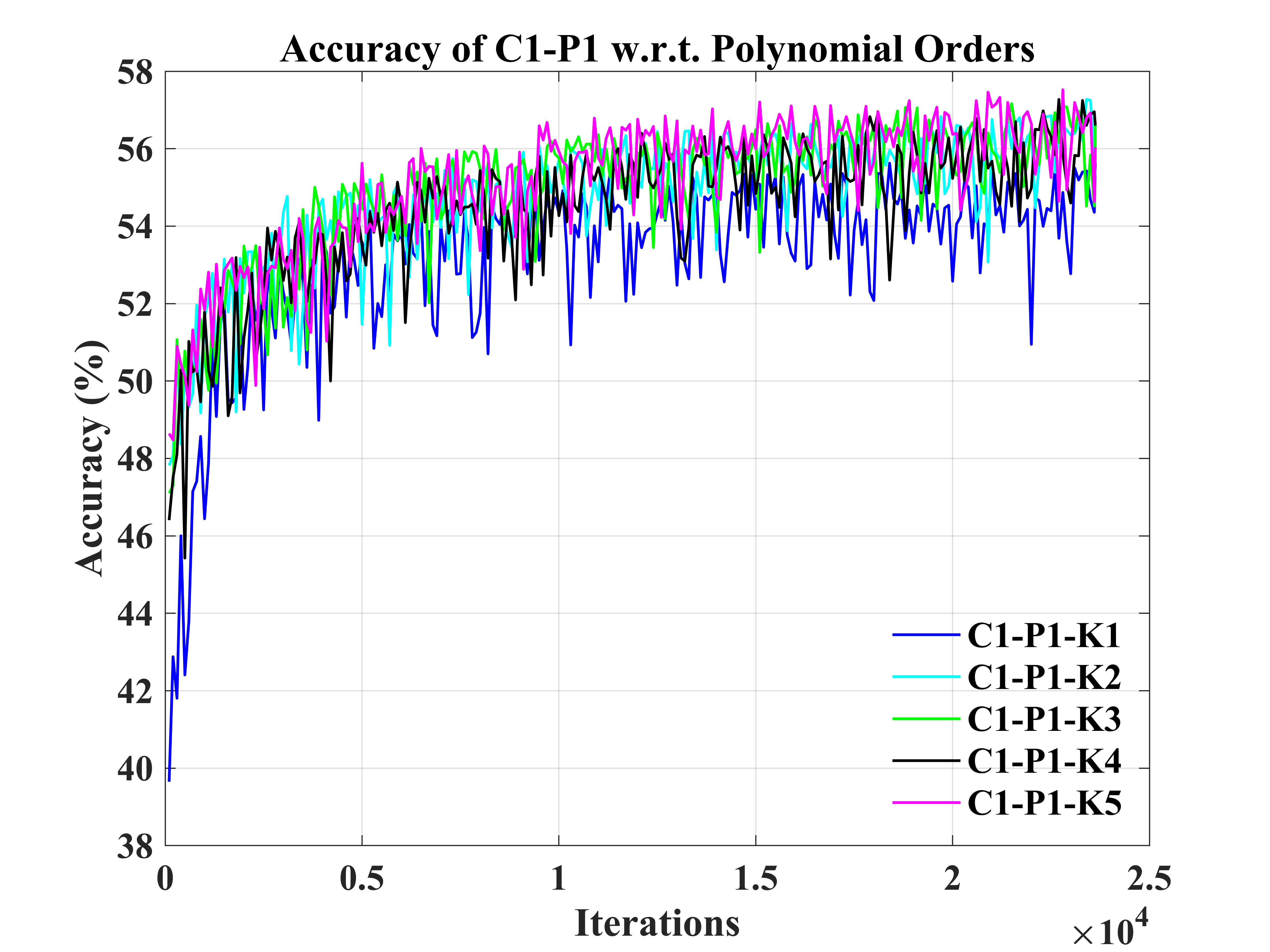}
			\subcaption{}
			\label{Accuracy of C1 P1}
		\end{minipage}
		\begin{minipage}[t]{.24\linewidth}
			\includegraphics[width=1.65in]{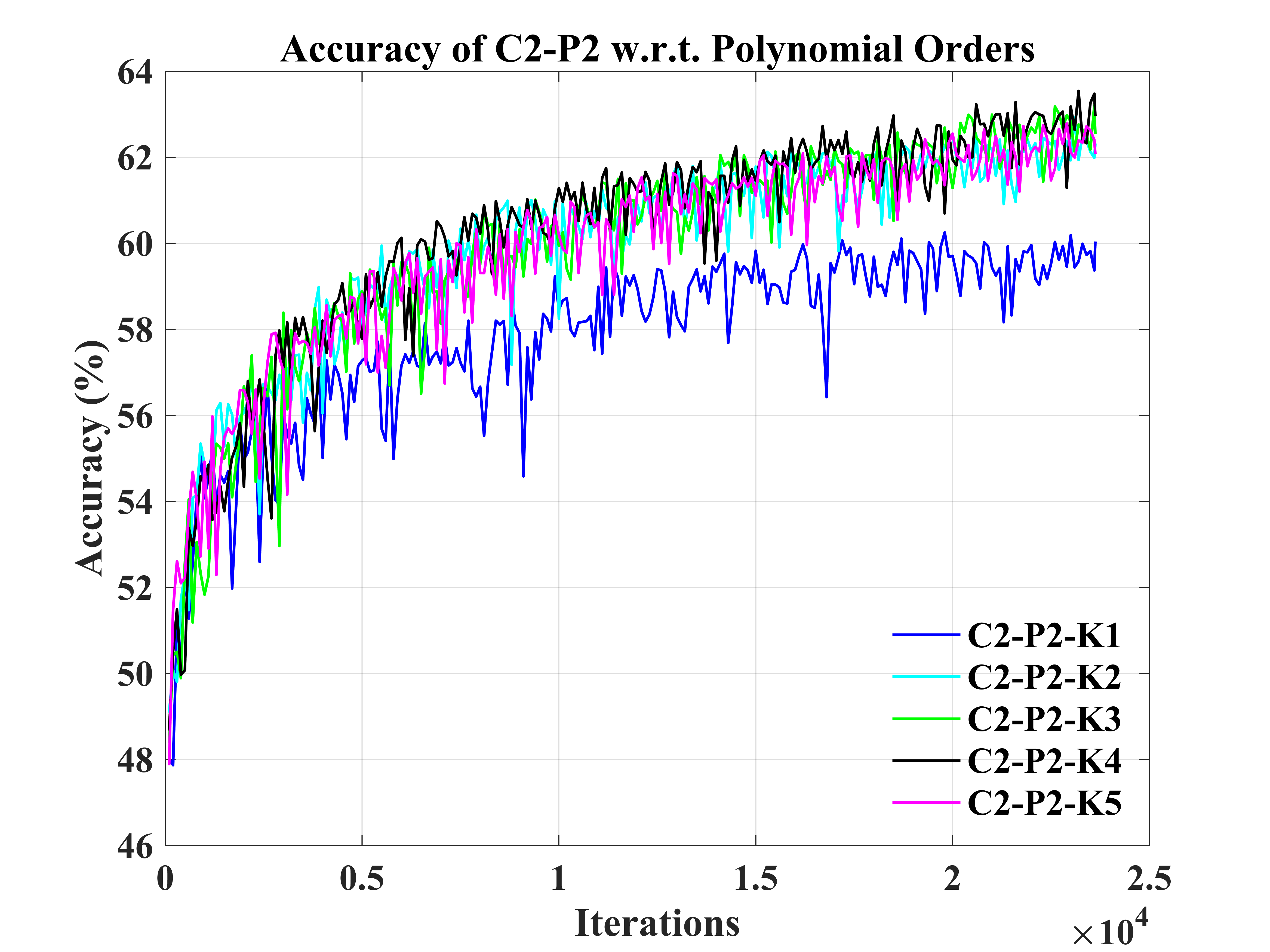}
			\subcaption{}
			\label{Accuracy of C2 P2}
		\end{minipage}
		\begin{minipage}[t]{.24\linewidth}
			\includegraphics[width=1.65in]{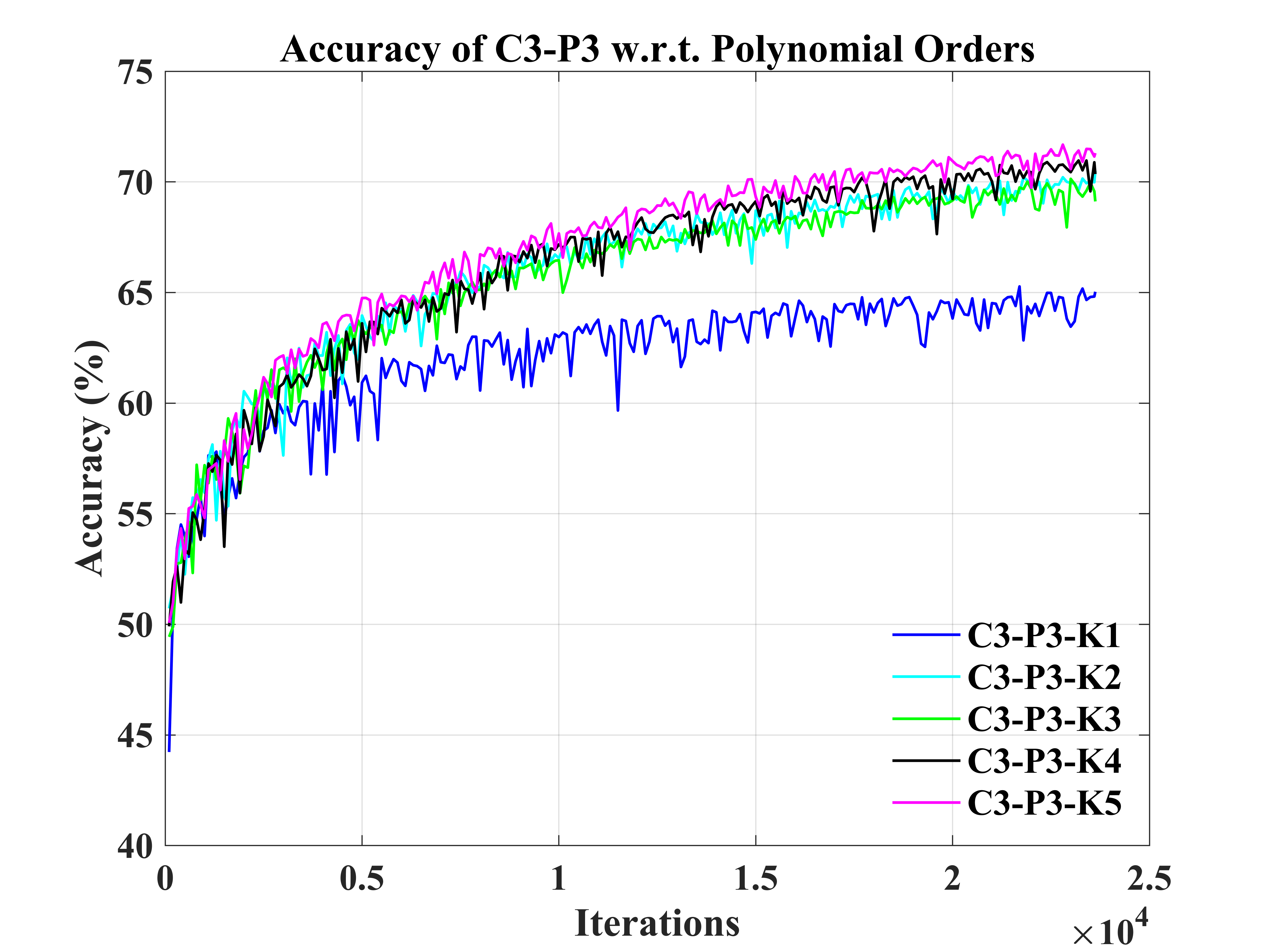}
			\subcaption{}
			\label{Accuracy of C3 P3}
		\end{minipage}
		\begin{minipage}[t]{.24\linewidth}
			\includegraphics[width=1.65in]{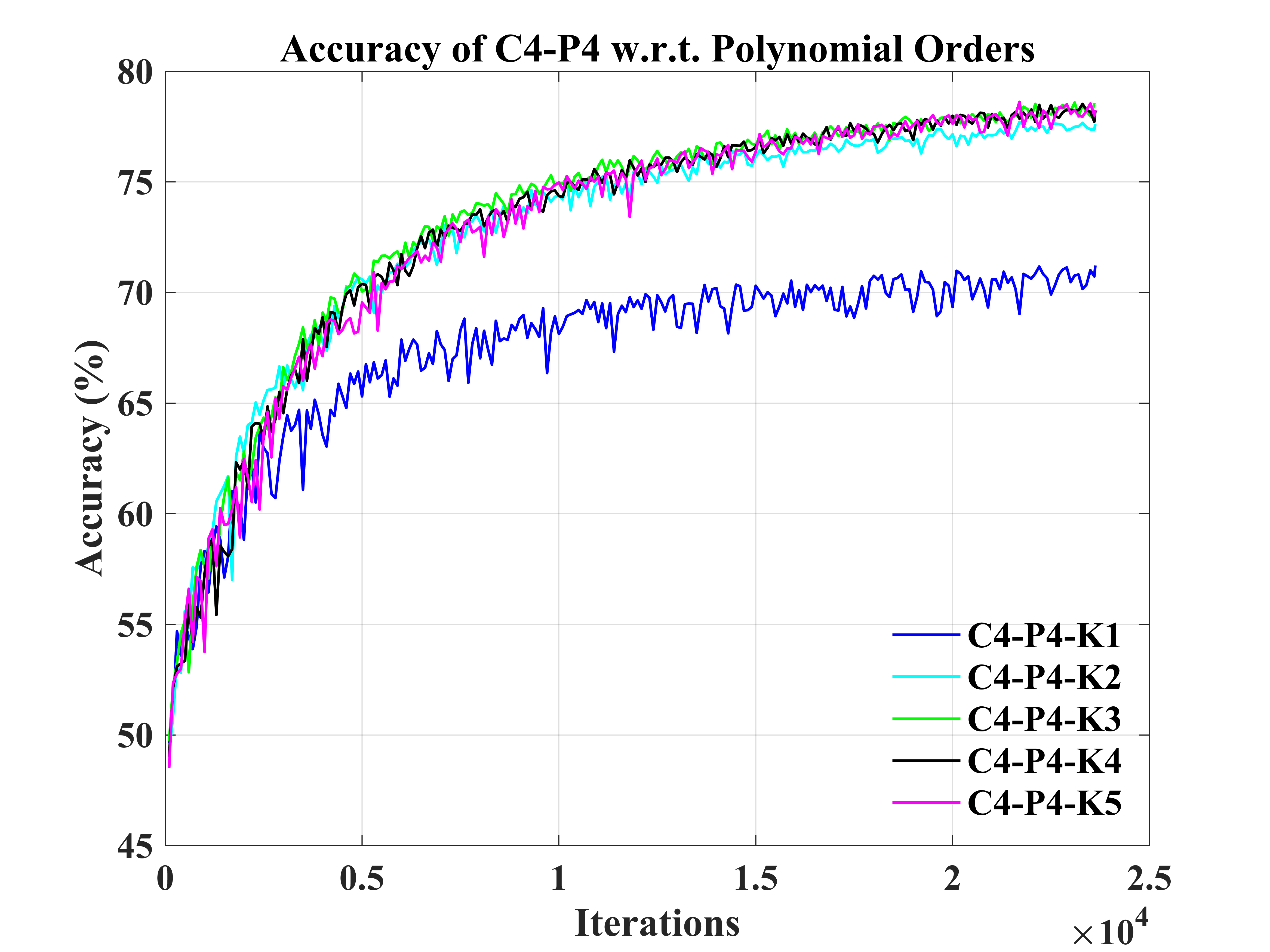}
			\subcaption{}
			\label{Accuracy of C4 P4}
		\end{minipage}
		\begin{minipage}[t]{.24\linewidth}
			\includegraphics[width=1.65in]{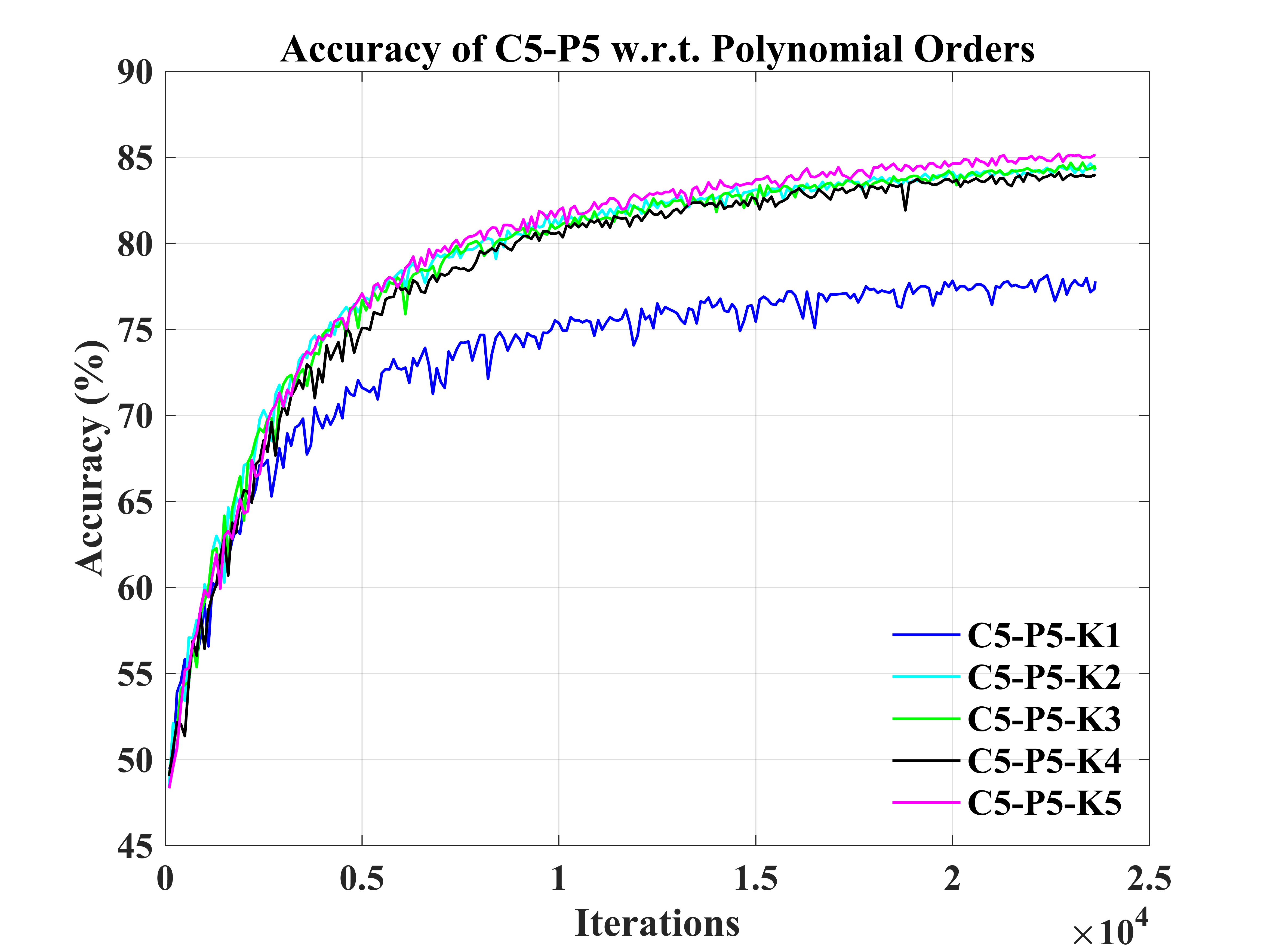}
			\subcaption{}
			\label{Accuracy of C5 P5}
		\end{minipage}
		\begin{minipage}[t]{.24\linewidth}
			\includegraphics[width=1.65in]{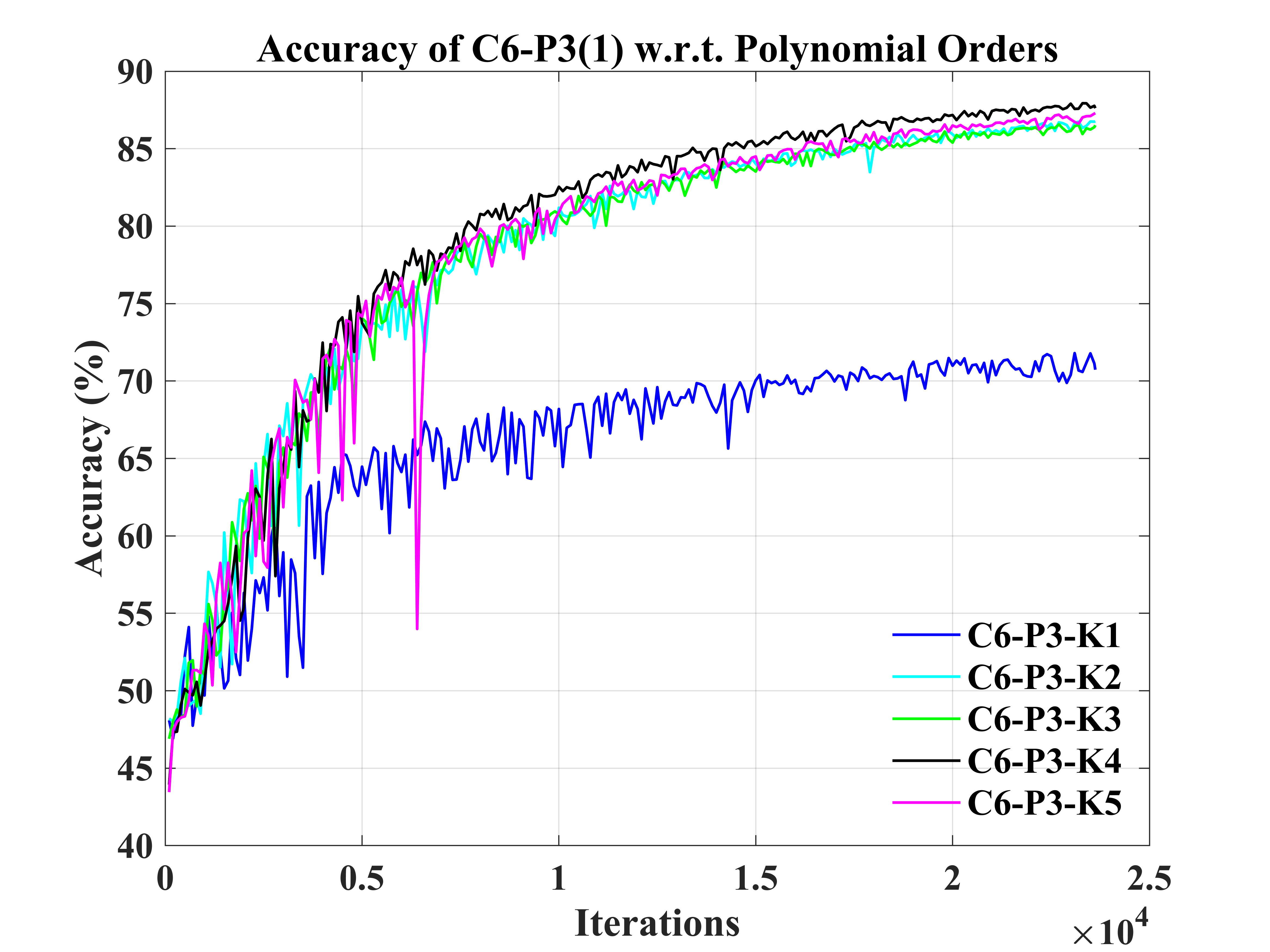}
			\subcaption{}
			\label{Accuracy of C6 P3-1}
		\end{minipage}
		\begin{minipage}[t]{.24\linewidth}
			\includegraphics[width=1.65in]{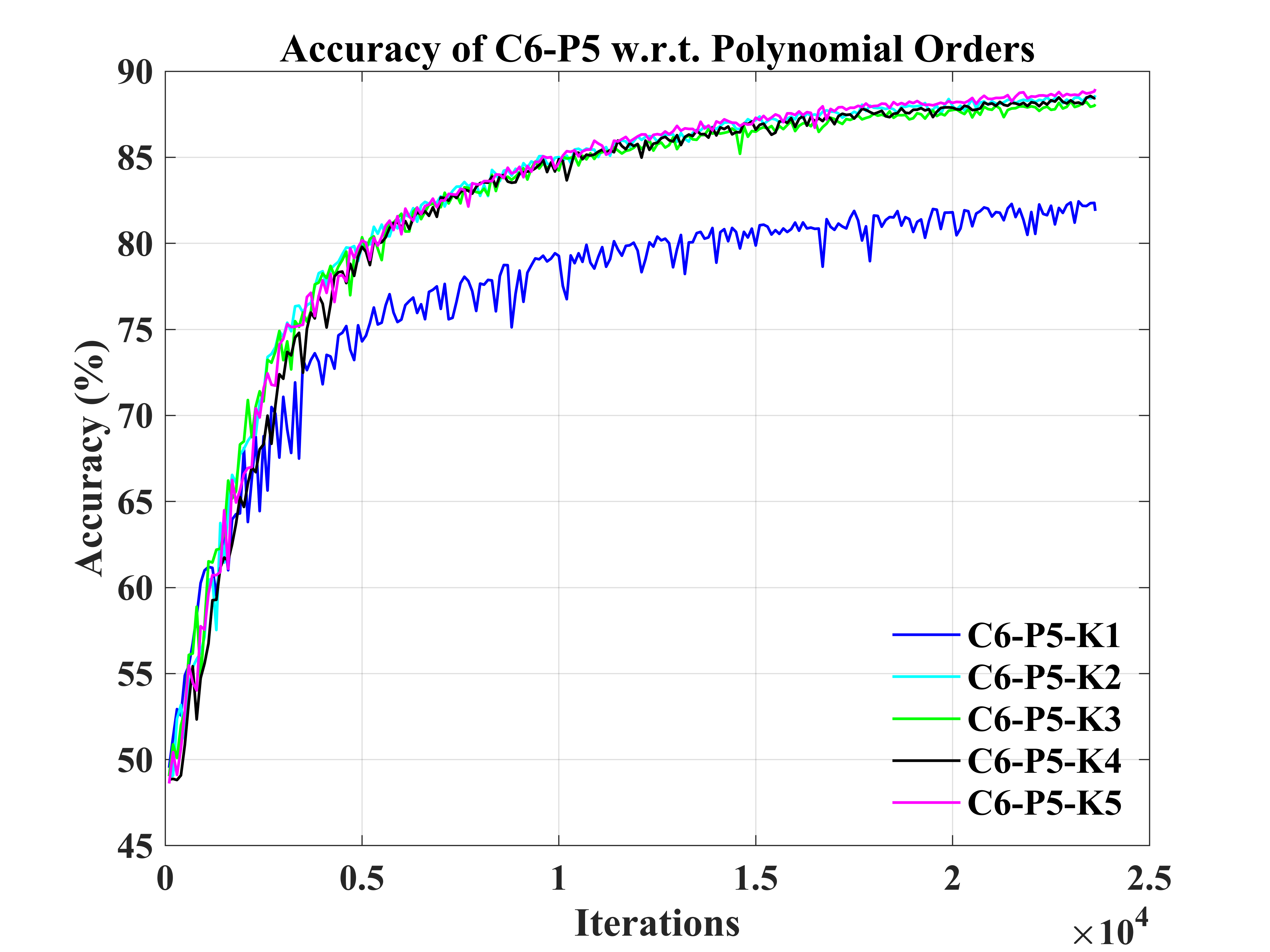}
			\subcaption{}
			\label{Accuracy of C6 P5}
		\end{minipage}
		\begin{minipage}[t]{.24\linewidth}
			\includegraphics[width=1.65in]{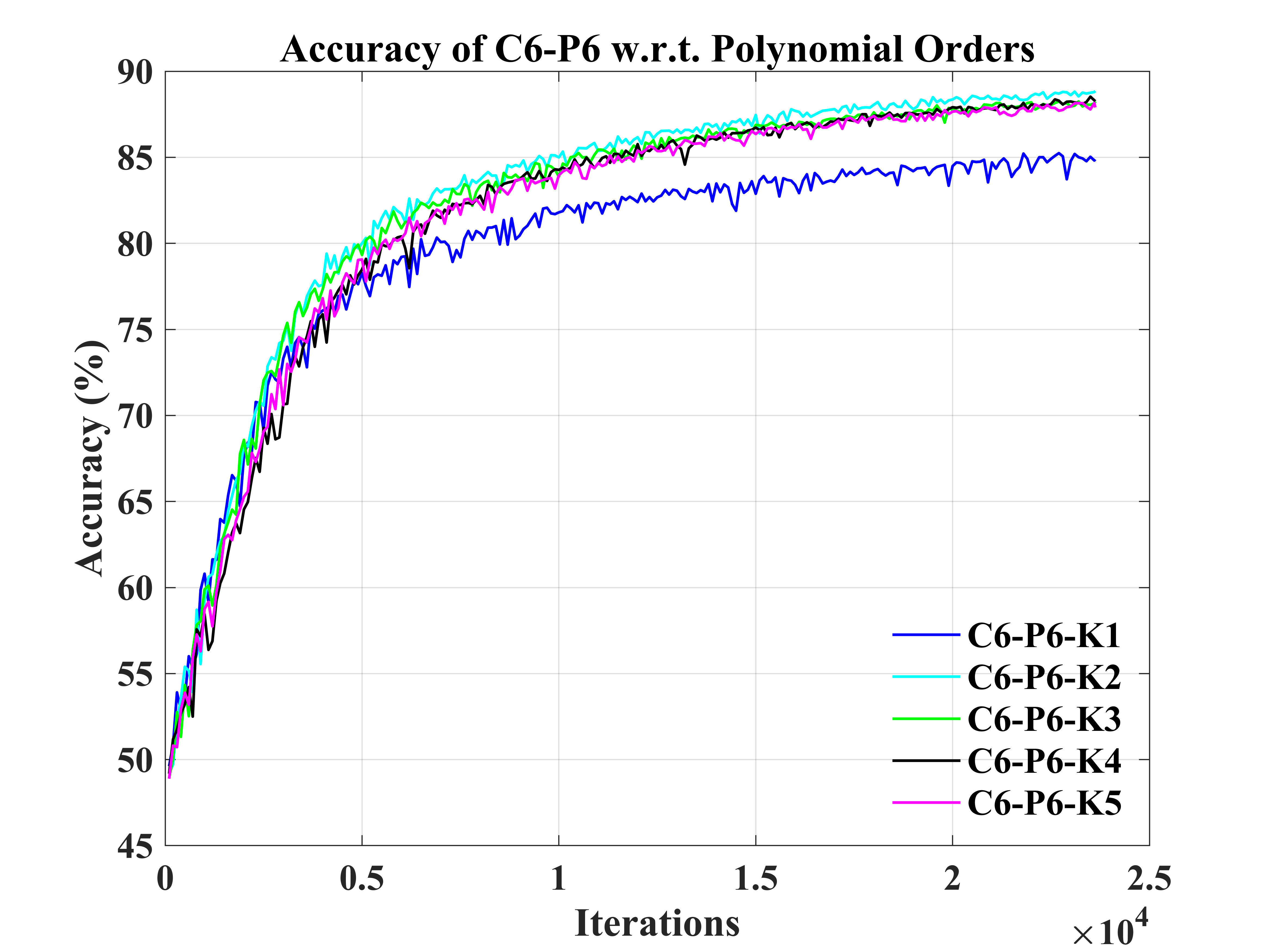}
			\subcaption{}
			\label{Accuracy of C6 P6}
		\end{minipage}
		\caption{Accuracy of some selected models regarding different polynomial approximation order. The models are selected from \reftab{Model framework in the experiments}. (a) Accuracy of the model C1-P1 (model 1). (b) Accuracy of the model C2-P2 (model 3). (c) Accuracy of the model C3-P3 (model 6). (d) Accuracy of the model C4-P4 (model 10). (e) Accuracy of the model C5-P5 (model 14). (f) Accuracy of the model C6-P3 (model 16). (g) Accuracy of the model C6-P5 (model 19). (h) Accuracy of the model C6-P6 (model 20).}
		\label{Accuracy w.r.t. polynomial orders}
	\end{figure*}
	
	\subsubsection{Graph Coarsening and Fast Pooling}\label{Graph Coarsening and Fast Pooling}
	Compared with the pooling operation in CNNs, on graphs, it involves nodes clustering and one-dimensional pooling. To carry out pooling to reduce dimensionality, the Graclus multilevel clustering algorithm is performed~\cite{dhillon2007weighted}. A greedy algorithm is employed to measure the consecutive coarser of the graph and minimize the objective of the spectral clustering~\cite{shi2000normalized}. 1) At each level, multiple numbers of the coarser graphs are given. 2) It picks an unmarked node $i$, and matches with its unmarked neighborhood $j$, which needs to maximize the local normalized cut $W_{i j}\left(1 / d_{i}+1 / d_{j}\right)$. $W_{i j}$ denotes the edge weight between the node $i$ and node $j$, and $d_{i}$ and $d_{j}$ are the distances between the coarsened node and the node $i$ and node $j$, respectively~\cite{dhillon2007weighted, shi2000normalized}. 3) It will mark the two matched nodes, and the sum of their weights will be the coarsened weight. 4) All the nodes will undergo the same procedure. Noticeably, at the coarsest level, the nodes will be arbitrarily ordered. Then, the ordered nodes will be propagated to the finest level. Finally, the graph signal is pooled in a one-dimensional manner~\cite{defferrard2016convolutional}. This algorithm cuts the number of nodes by two between two levels.
	
	\subsection{Model Initialization}\label{Model Initialization}
	A novel structure of the GCNs is introduced to classify EEG MI tasks. Based on the absolute Pearson’s matrix of overall signals, the graph Laplacian is built up to represent the topological relationship of EEG electrodes. The graph convolutional layers learn the generalized features. Built on a maximum of $\rm \log_2{N}$ graph pooling layers regarding $\rm N$ EEG channels, the pooling operation reduces dimensionality, and the FC softmax layer derives the final prediction. With regard to 64-channel international 10-10 EEG system, the maximum number of pooling layers is six. The implementation details are listed in~\reftab{Implementation details for proposed GCNs-Net}, where $\rm N$ denotes the input size of the EEG signals, ${\rm F}_i\in\left[{\rm F}_1, {\rm F}_2, {\rm F}_3, {\rm F}_4, {\rm F}_5, {\rm F}_6\right]$ donates the number of filters at the $i^{th}$ graph convolutional layer, $\rm K$ denotes the polynomial order for filters, and $\rm O$ is the number of MI tasks.
	
	The hyperparameters in our work (\eg, learning rate, dropout rate, and weight decay rate) are mainly empirically chosen over all the experiments, which are not task-oriented tuned. The network parameters,~\ie, weights and biases, are updated by the Adam iterative solver~\cite{kingma2015adam} with a 0.01 learning rate. Biases are applied to every node of the graph. The batch size is 1,024 to maximize the usage of GPU resources. For FC layers in~\refsec{Novel Deep Learning Framework of the GCNs}, a 50\% dropout rate is applied~\cite{srivastava2014dropout}. The batch normalization (BN) is employed for graph convolutions. The BN normalizes the input graph signals by subtracting the mean and dividing the standard deviation of the mini-batch. Then, it scales and shifts the normalized signals to align with the original distribution. It not only alleviates the problem of internal covariate shift but also prevents gradient vanishing~\cite{ioffe2015batch}. The non-linear Smooth Rectified Linear Unit (Softplus) activation function is applied to the graph convolutional layers and FC layers to prevent gradient vanishing~\cite{7280459}, where $\boldsymbol{\rm x}$ are the input signals.
	\begin{equation}
		\mathrm{Softplus}(\boldsymbol{\rm x})=\log\left(1+\mathrm{exp}\left(\boldsymbol{\rm x}\right)\right).
	\end{equation}
	
	The softmax function is utilized to derive the final prediction. 
	\begin{equation}\label{Output}
		\mathrm{Softmax}(\hat{y_i})=\frac{\mathrm{exp}\left(\hat{y_i}\right)}{\sum\limits_{i=1}^{\rm O} \mathrm{exp}\left(\hat{y_i}\right)},
	\end{equation}
	where $\hat{y_i}\in[\hat{y_1}, \cdots, \hat{y_{\rm O}}]$ is the predicted probability of the $i^{th}$ MI task. In \refequ{loss}, the cross-entropy loss with the L2 regularization (weight decay) is employed as the loss function. Meanwhile, the weight decay rate $\rho$ is set to $1\times10^{-2}$.
	\begin{equation}
		\label{loss}
		\mathrm{Loss}=-\sum_{i=1}^{\rm O} y_{i} \log \left(\hat{y}_{i}\right) +\frac{\rho}{2N_{P}}\|{\boldsymbol{\rm W}}\|^{2}.
	\end{equation}
	$y_{i}$, $\boldsymbol{\rm W}$, and $N_{P}$ are the corresponding MI task, network parameters, and the number of parameters, respectively.
	
	\begin{table*}[!ht]
		\centering
		\caption{Model performance comparisons by changing the number of FC layers and the number of filters}
			\begin{tabular}{cccccccccc}
				\toprule
				Model & \tabincell{c}{Num. of \\ Filters} & \tabincell{c}{Num. of \\ FC Layers} & \tabincell{c}{Num. of \\ Neurons at FC Layer} & Model Framework & Accuracy \\ \midrule
				Base & 16, 32, 64, 128, 256, 512 & 1 & 4 & (C-P)$\times$6-S & 88.85\%\\ 
				1 & 32, 64, 128, 256, 512, 1024 & 1 & 4 & (C-P)$\times$6-S & 90.60\% \\ 
				2 & 64, 128, 256, 512, 1024, 1536 & 1 & 4 & (C-P)$\times$6-S & 90.89\% \\ 
				3 & 16, 32, 64, 128, 256, 512 & 2 & 64, 4 & (C-P)$\times$6-F-S & 88.08\% \\ 
				4 & 16, 32, 64, 128, 256, 512 & 2 & 512, 4 & (C-P)$\times$6-F-S & 88.64\% \\ 
				5 & 16, 32, 64, 128, 256, 512 & 3 & 512, 64, 4 & (C-P)$\times$6-F$\times$2-S & 87.36\% \\ 
				6 & 16, 32, 64, 128, 256, 512 & 3 & 512, 256, 4 & (C-P)$\times$6-F$\times$2-S & 88.35\% \\ 
				7 & 32, 64, 128, 256, 512, 1024 & 3 & 512, 64, 4 & (C-P)$\times$6-F$\times$3-S & 90.45\% \\ \bottomrule
			\end{tabular}
		\label{Model Framework by Changing Num of FC Layers and filter's Amount}
	\end{table*}
	
	\begin{figure}[!ht]
		\centering
		\begin{minipage}[t]{.48\linewidth}
			\includegraphics[width=1.8in]{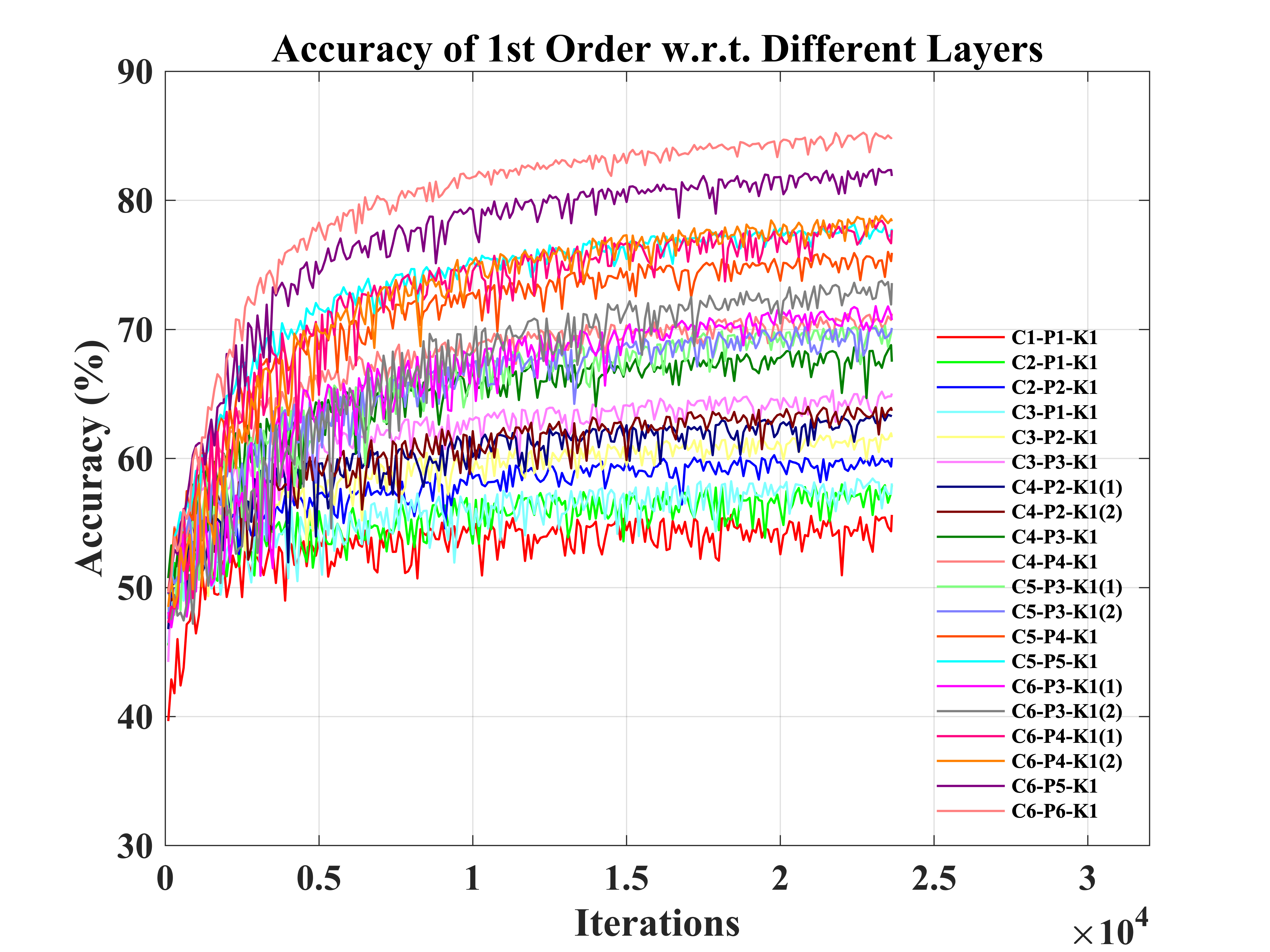}
			\subcaption{}
			\label{Accuracy of 1st Order}
		\end{minipage}
		\begin{minipage}[t]{.48\linewidth}
			\includegraphics[width=1.8in]{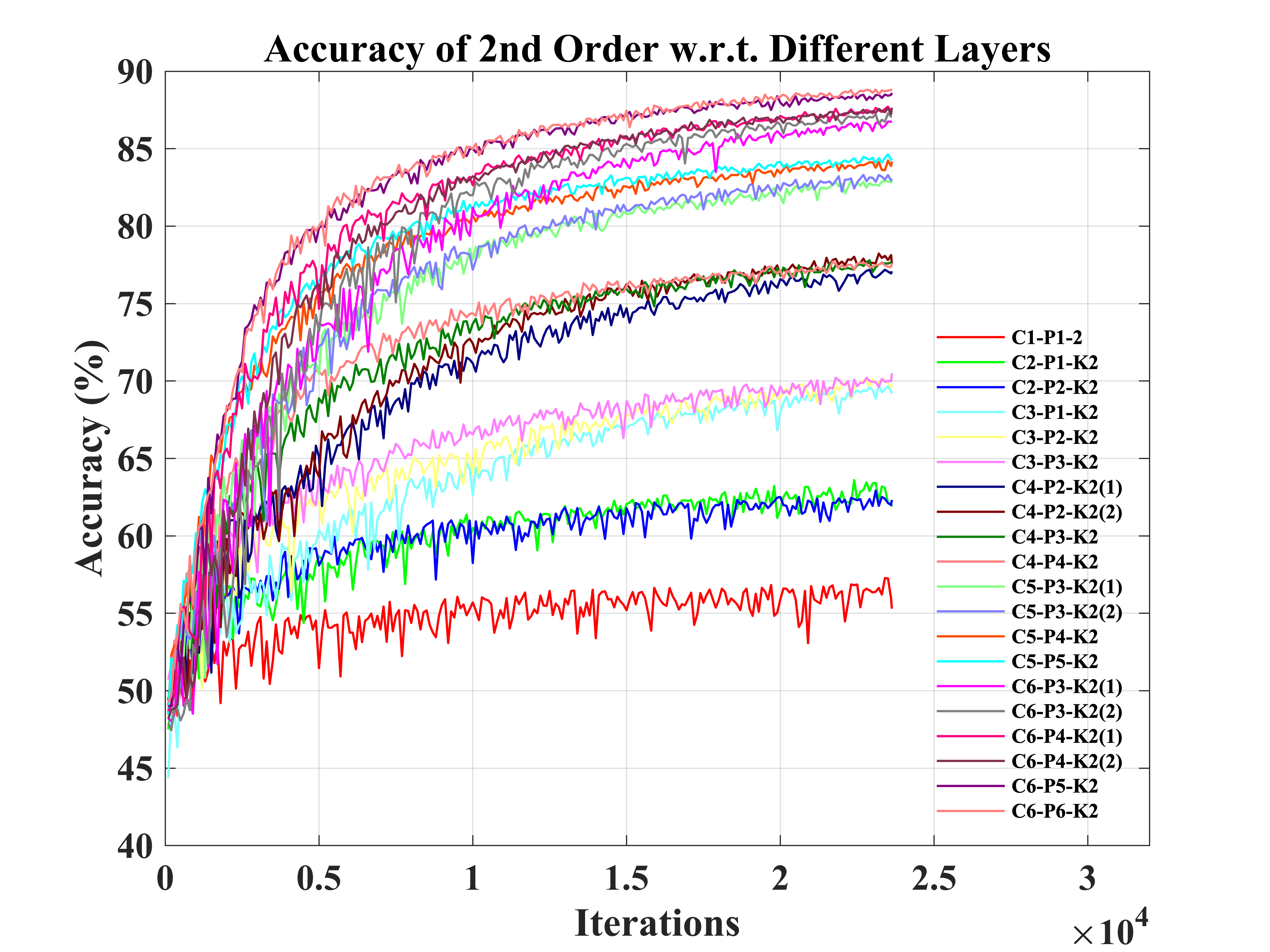}
			\subcaption{}
			\label{Accuracy of 2nd Order}
		\end{minipage}
		\begin{minipage}[t]{.48\linewidth}
			\includegraphics[width=1.8in]{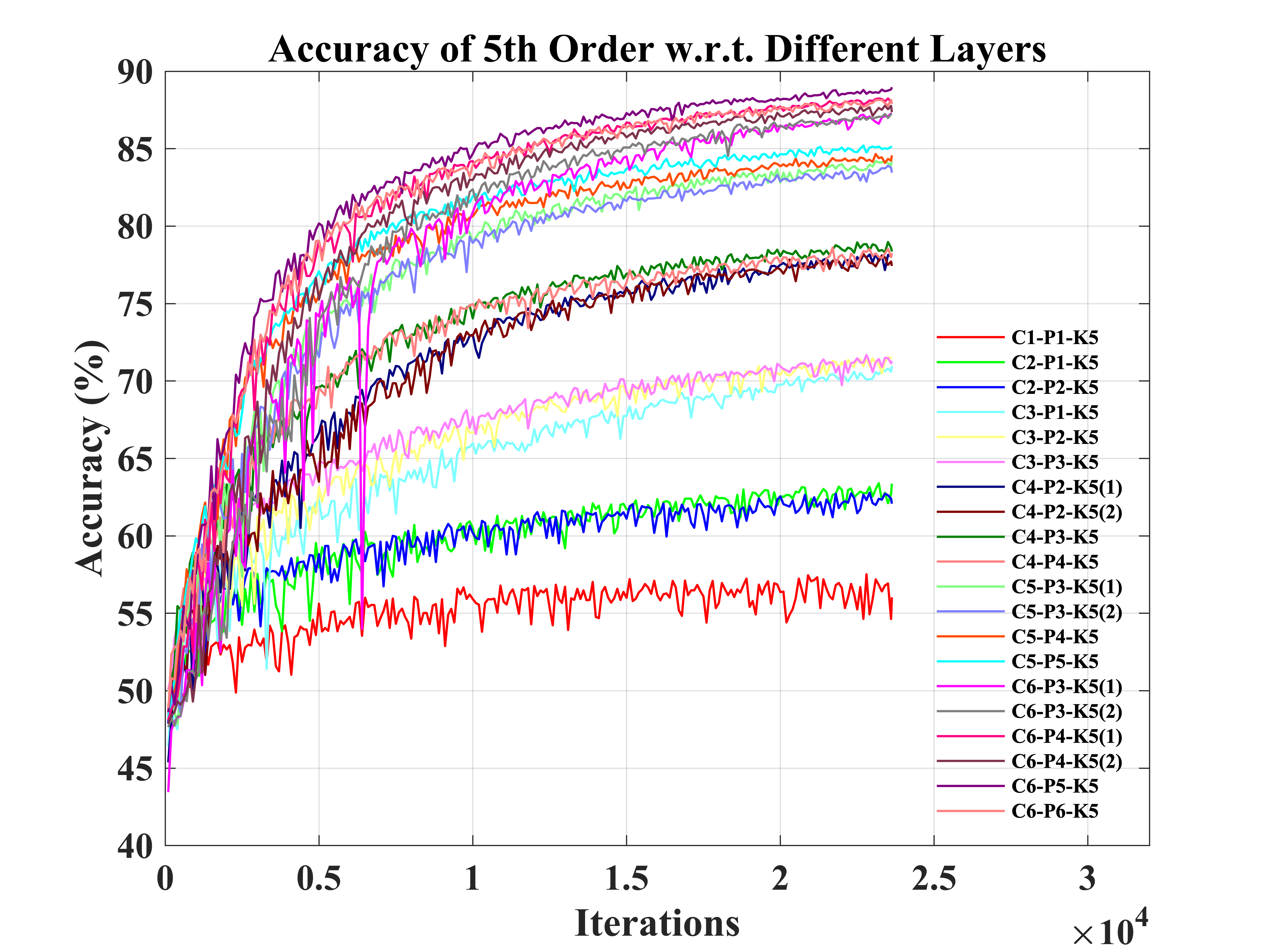}
			\subcaption{}
			\label{Accuracy of 5th Order}
		\end{minipage}
		\begin{minipage}[t]{.48\linewidth}
			\includegraphics[width=1.8in]{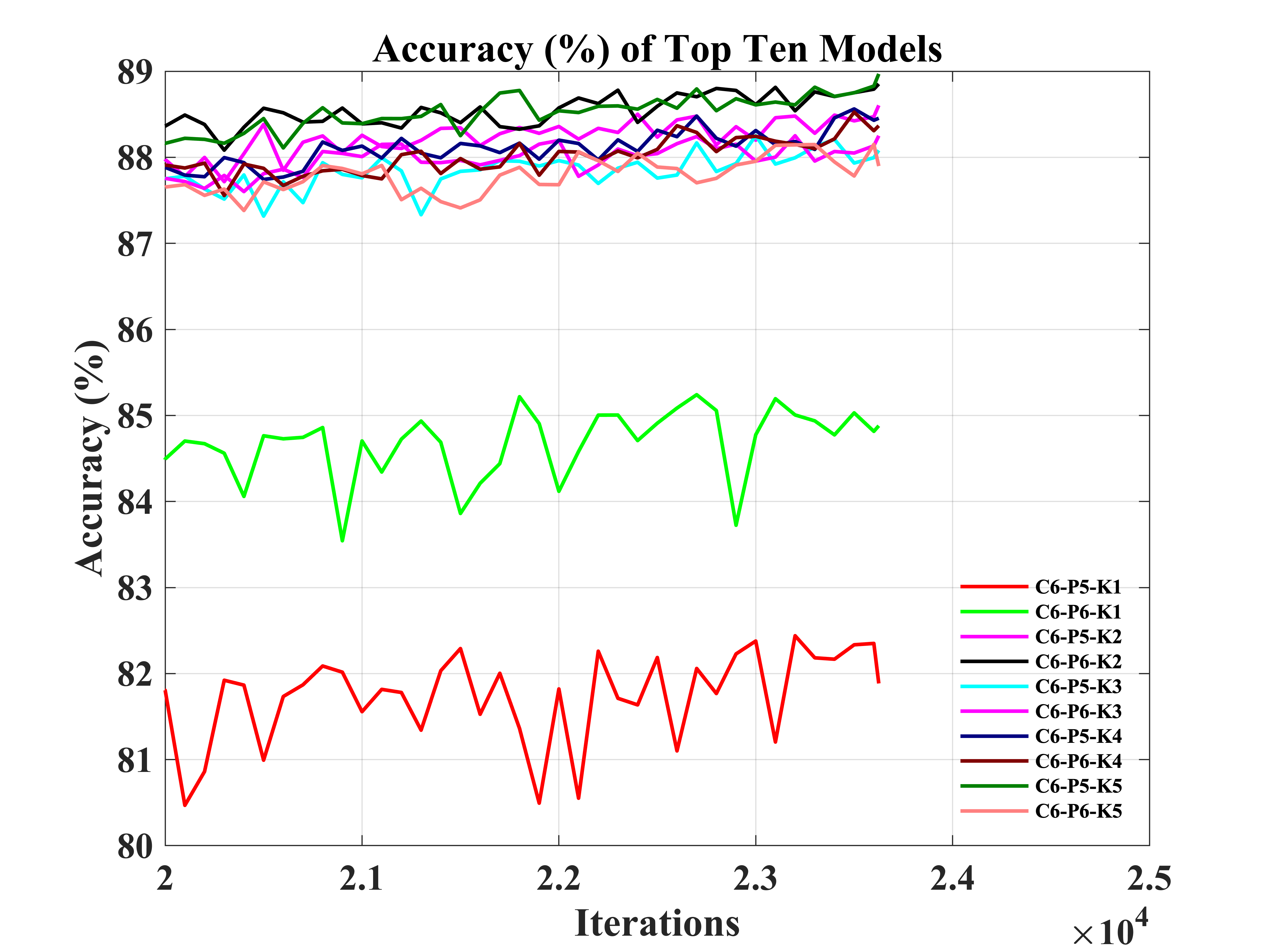}
			\subcaption{}
			\label{Top Ten}
		\end{minipage}
		\caption{Accuracy of different models while applying the same polynomial order. (1) Accuracy of different models regarding the $1^{st}$ order Chebyshev polynomial. (b) Accuracy of different models regarding the $2^{nd}$ order Chebyshev polynomial. (c) Accuracy of different models regarding the $5^{th}$ order Chebyshev polynomial. (d) Accuracy of the top ten models.}
		\label{Accuracy of different models}
	\end{figure}
	
	\subsection{Evaluation Metrics}\label{Evaluation Metrics}
	To evaluate performance, multiple metrics are adopted, including accuracy, Cohen's Kappa coefficient (Kappa)~\cite{cohen1960coefficient}, single class accuracy on each task, Macro-averaged precision, recall, F1-score, Receiver Operating Characteristic Curve (ROC curve), and the Area Under ROC Curve (AUC). Besides, to validate whether the performance difference between methods is statistically significant, the pair-wise $t$-test is applied. In this work, the significance level of the $t$-test,~\ie, the $p$-value, is set to 0.05~\cite{zhang2018learning}.
	
	\section{Results and Discussion}\label{Results}
	\subsection{A Novel Deep Learning Framework of the GCNs}\label{Novel Deep Learning Framework of the GCNs}
	To explore an optimal model for the GCNs-Net, as detailed in~\reftab{Model framework in the experiments}, the decoding performance of multiple structures is investigated by changing some network hyperparameters, such as the number of graph convolutional (Conv) and pooling layers, the polynomial order of the Chebyshev polynomials for filters, and the number of convolutional filters. C denotes a graph convolutional layer, P denotes a graph pooling layer, F denotes an FC layer, and S denotes a softmax layer. The PhysioNet Dataset is used to compare the performance of different architectures, as it contains the largest number of participants in the field of EEG MI. The amount of data makes it particularly well suited for training DL models. The dataset of 20 subjects (S$_1$$\sim$S$_{20}$) with 64-channel 1,075,200 samples (640 time points $\times$ 84 trials $\times$ 20 subjects) is utilized to train and evaluate different architectures.
	
	First of all, while holding the number of graph Conv and pooling layers, experiments are carried out by changing the Chebyshev polynomial order from $1^{st}$ to $5^{th}$ as described in~\reftab{Model framework in the experiments}.~\reffig{Accuracy of C1 P1} displays that when there is only one graph Conv layer followed by a graph pooling layer, the order of the Chebyshev does not make a difference. The accuracy regarding each order is less than 58\%. They overlap with each other and fluctuate during training. Additionally, when the number of graph Conv layers is greater than one, the accuracy of the model with the $1^{st}$ order approximation ascends. In the later epochs, it enters a period of dormancy. The accuracy of models with the $1^{st}$ order witnesses a rugged and abrupt ascent when there are more graph Conv layers. Apart from the number of graph Conv layers, the accuracies of models move upward smoothly with the increasing number of pooling layers.
	
	As illustrated in~\reffig{Accuracy w.r.t. polynomial orders}, the accuracy regarding models with the $1^{st}$ polynomial order is unsatisfactory. By contrast, the accuracies of models regarding the $2^{nd}$ to $5^{th}$ polynomial orders are making a different climb. But they almost overlap and parallel with each other during training. It indicates that when the order of polynomial approximation is greater than one, there is a minor impact on the EEG MI decoding. As a result, $2^{nd}$ order Chebyshev approximation for filters is employed to not only achieve a superior performance but also reduce the model complexity.
	
	Besides, the impacts of performance by changing the number of graph Conv and pooling layers are ablated at a specific polynomial approximation order.~\reffig{Accuracy of different models} demonstrates that when the number of graph Conv layers increases, the accuracies take a steep climb. Notably, as for the $2^{nd}$ polynomial order in~\reffig{Accuracy of 2nd Order}, it illustrates that the number of Conv layers does affect performance. While applying a deeper model, including extra Conv layers, features can be better extracted from EEG signals. Meanwhile, the effects of the number of graph pooling layers are also investigated. The number of pooling layers promotes and enhances the decoding performance, but with a modest increment. As detailed in~\reffig{Top Ten}, when the polynomial order is 2$^{nd}$, the accuracies are 88.60\% (Model C6-P5-K2) and 88.85\% (Model C6-P6-K2), respectively.
	
	Furthermore, based on the optimal C6-P6-K2 model which contains six graph Conv layers with the $2^{nd}$ polynomial order for filters and six graph pooling layers, the influence on performances by changing the number of convolutional filters at every Conv layer and the number of FC layers is also explored in~\reftab{Model Framework by Changing Num of FC Layers and filter's Amount}.
	
	\begin{figure}[h]
		\centering
		\begin{minipage}[t]{.48\linewidth}
			\includegraphics[width=1.8in]{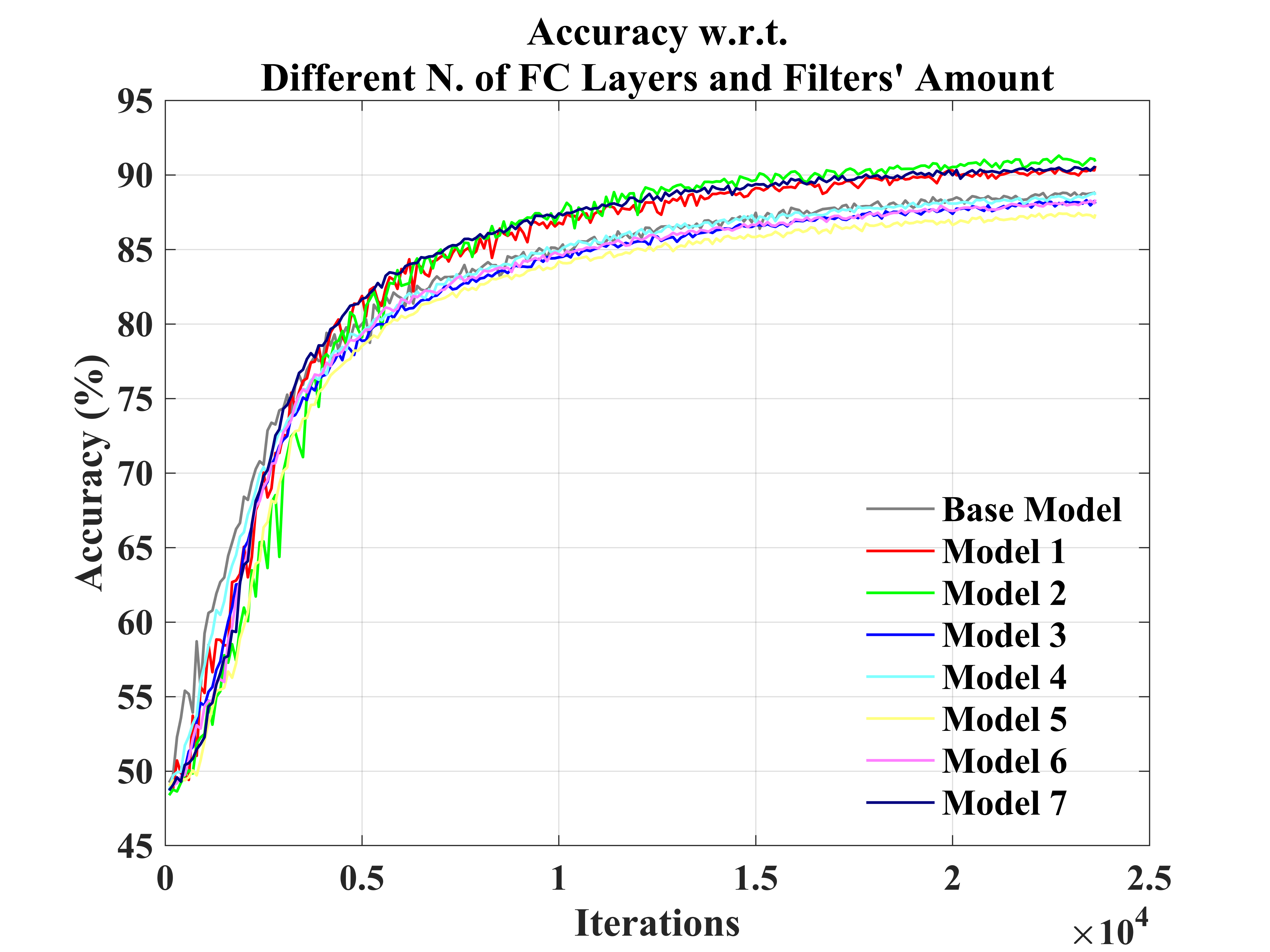}
			\subcaption{}
		\end{minipage}
		\begin{minipage}[t]{.48\linewidth}
			\includegraphics[width=1.8in]{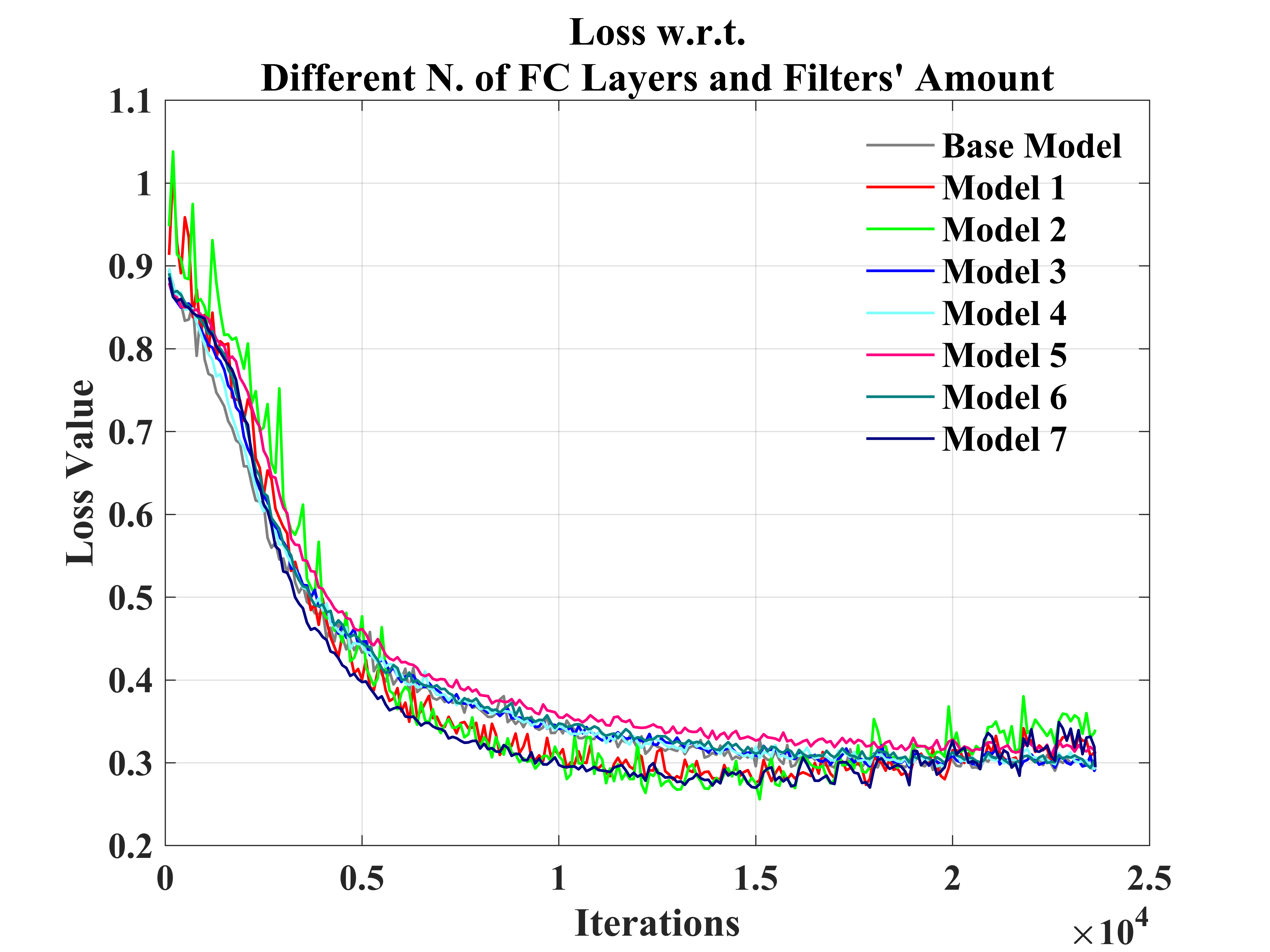}
			\subcaption{}
			\label{loss of filters}
		\end{minipage}
		\caption{Accuracy and loss of models with different numbers of FC layers and filters. (a) Accuracy of models with different numbers of FC layers and filters. (b) Loss of models with different numbers of FC layers and filters.}
		\label{Global_Average_Accuracy_of_FC_Filters}
	\end{figure}
	\begin{figure*}[ht]
		\centering
		\begin{minipage}[t]{.24\linewidth}
			\includegraphics[width=1.65in]{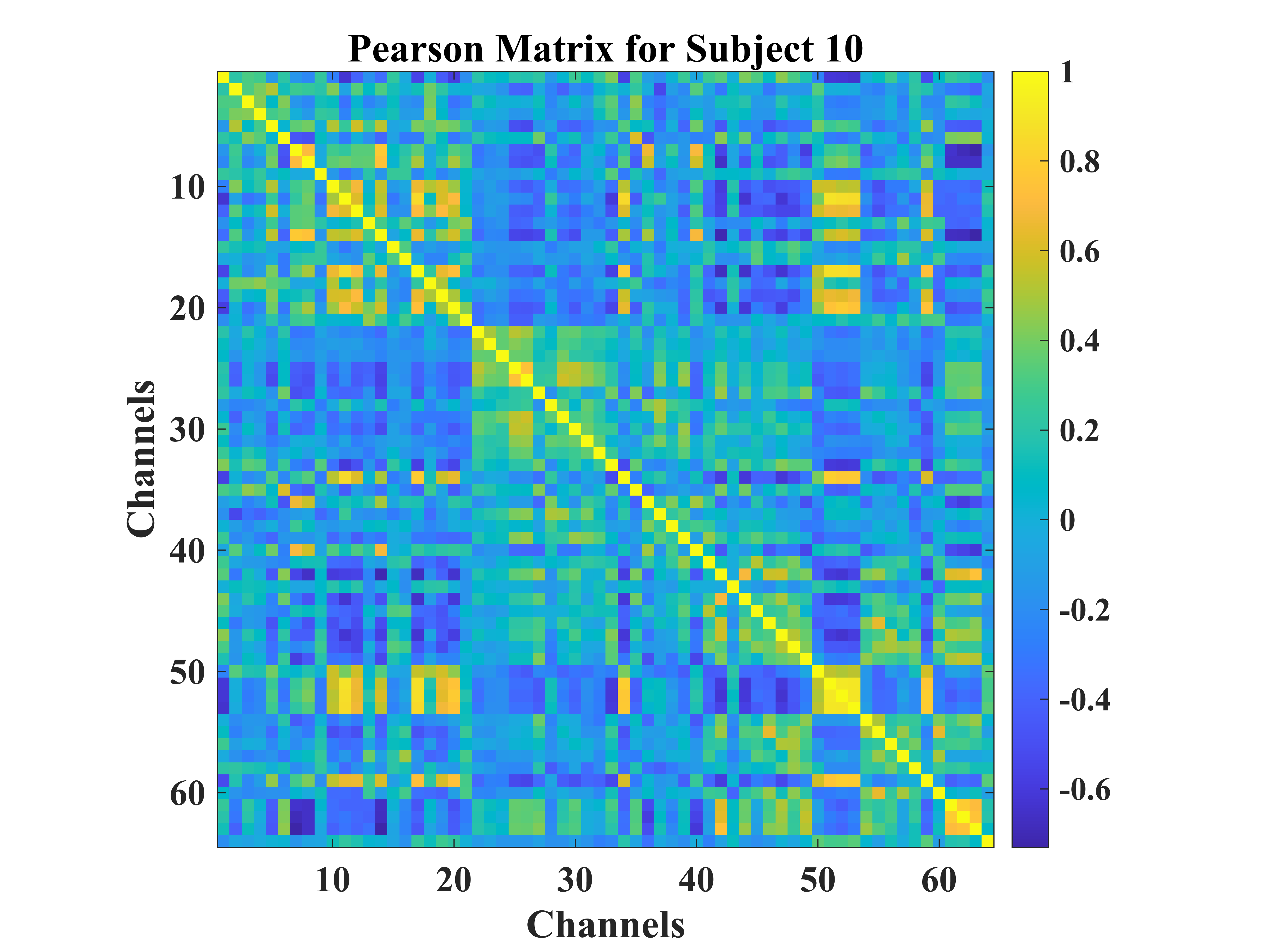}
			\subcaption{}
			\label{Pearson_matrix_for_Subject_10}
		\end{minipage}
		\begin{minipage}[t]{.24\linewidth}
			\includegraphics[width=1.65in]{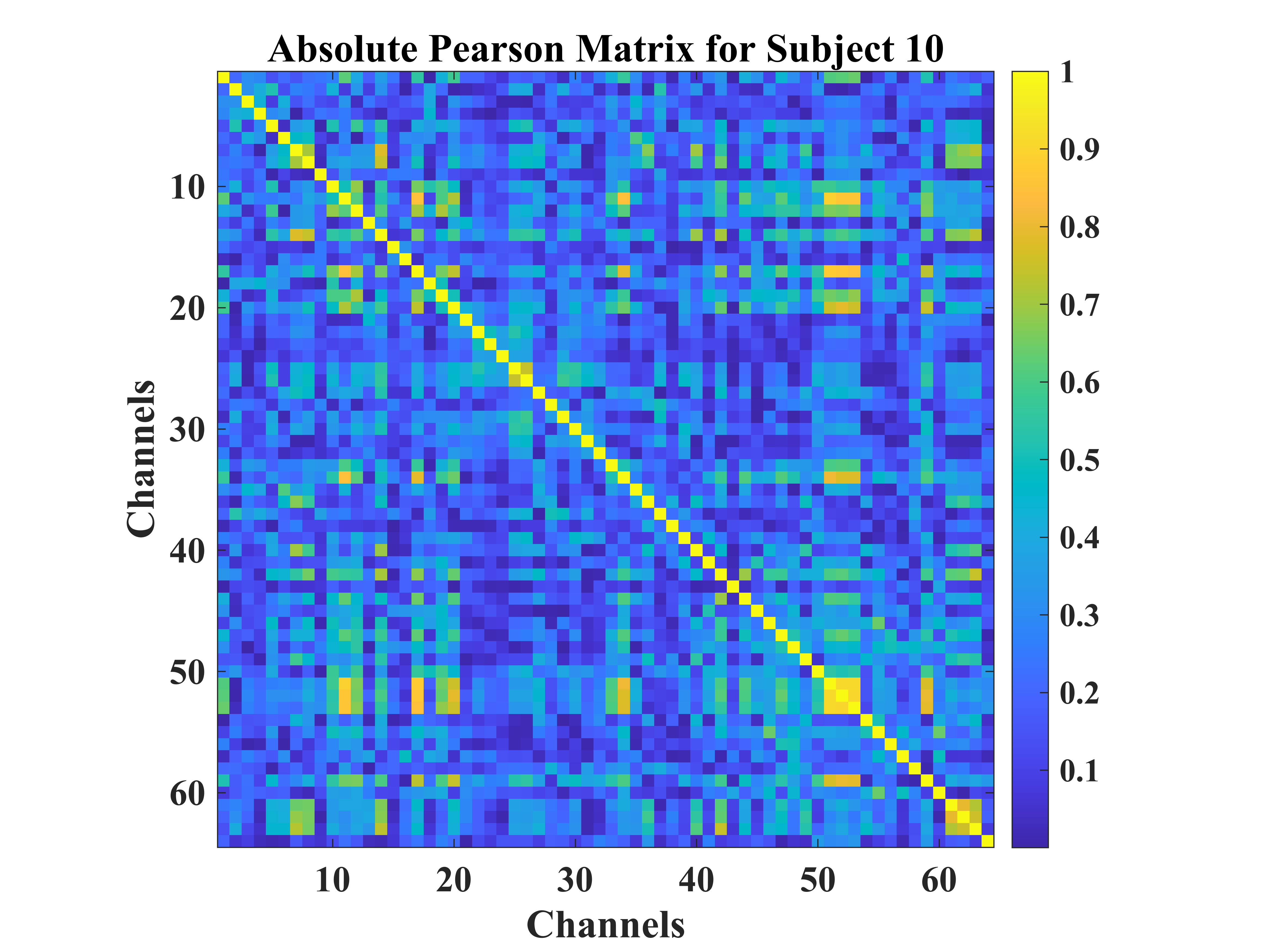}
			\subcaption{}
			\label{Absolute_Pearson_matrix_for_Subject_10}
		\end{minipage}
		\begin{minipage}[t]{.24\linewidth}
			\includegraphics[width=1.65in]{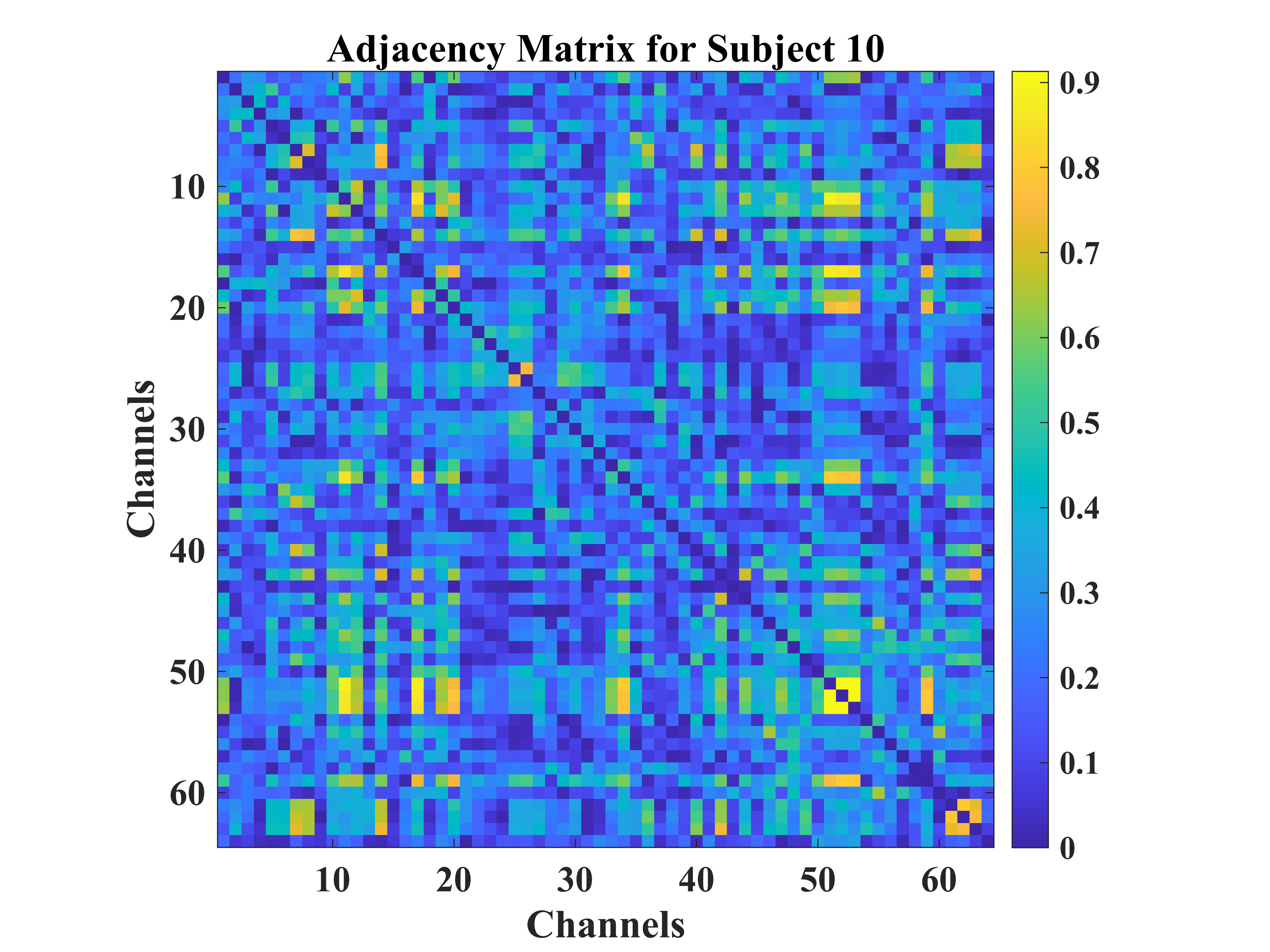}
			\subcaption{}
			\label{Adjacency_Matrix_for_Subject_10}
		\end{minipage}
		\begin{minipage}[t]{.24\linewidth}
			\includegraphics[width=1.65in]{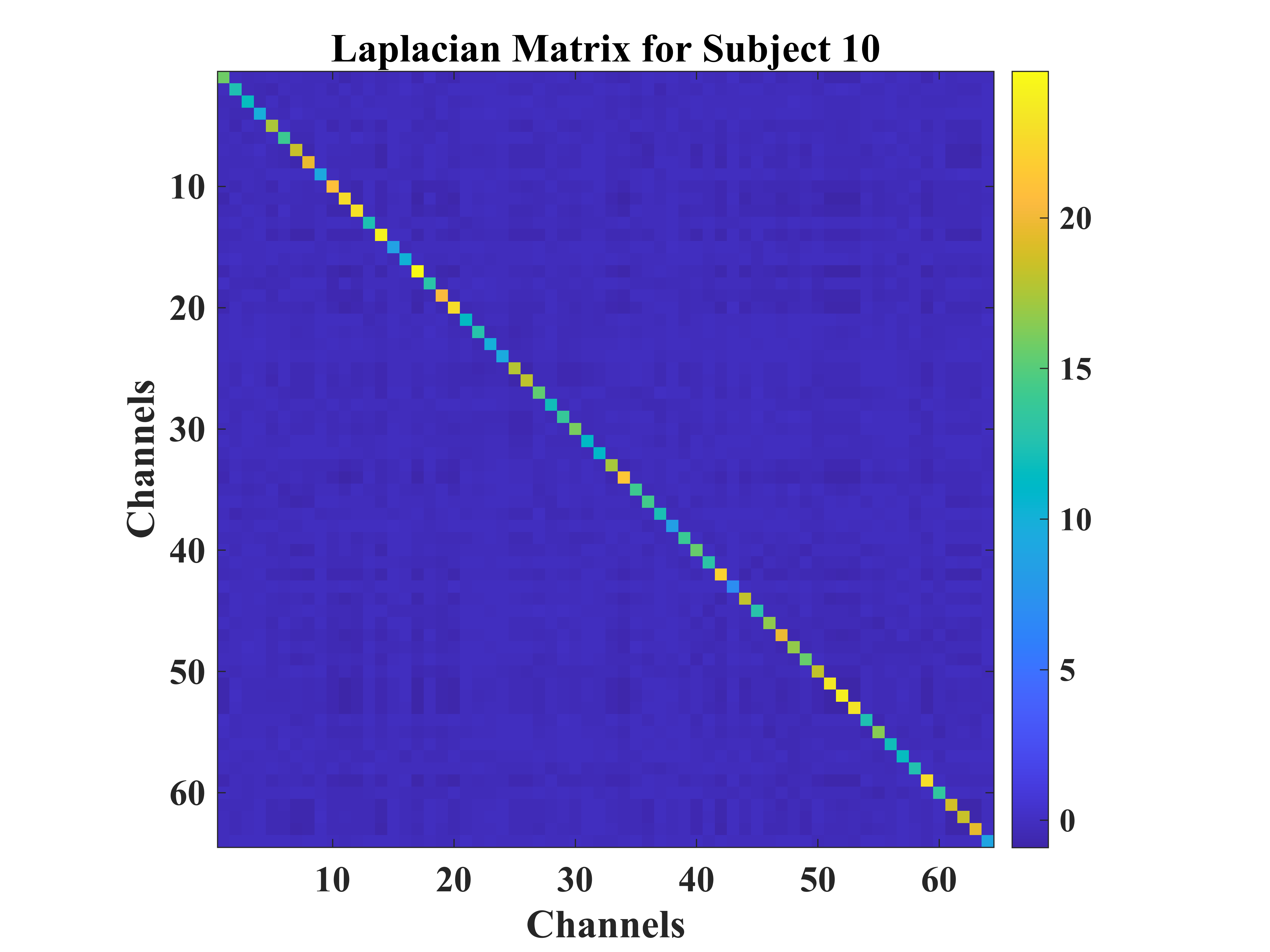}
			\subcaption{}
			\label{Laplacian_Matrix_for_Subject_10}
		\end{minipage}
		\begin{minipage}[t]{.24\linewidth}
			\includegraphics[width=1.65in]{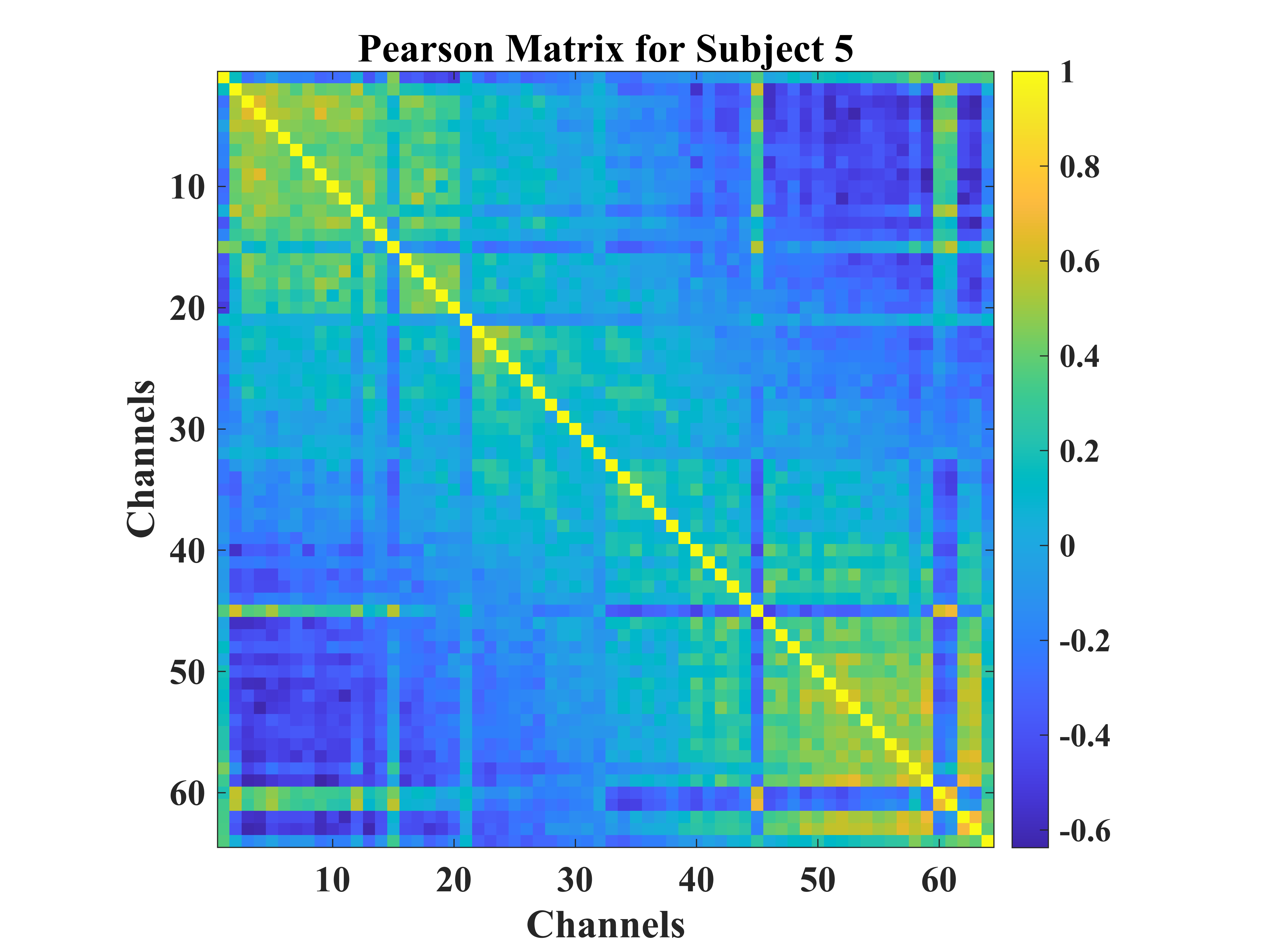}
			\subcaption{}
			\label{Pearson_matrix_for_Subject_5}
		\end{minipage}
		\begin{minipage}[t]{.24\linewidth}
			\includegraphics[width=1.65in]{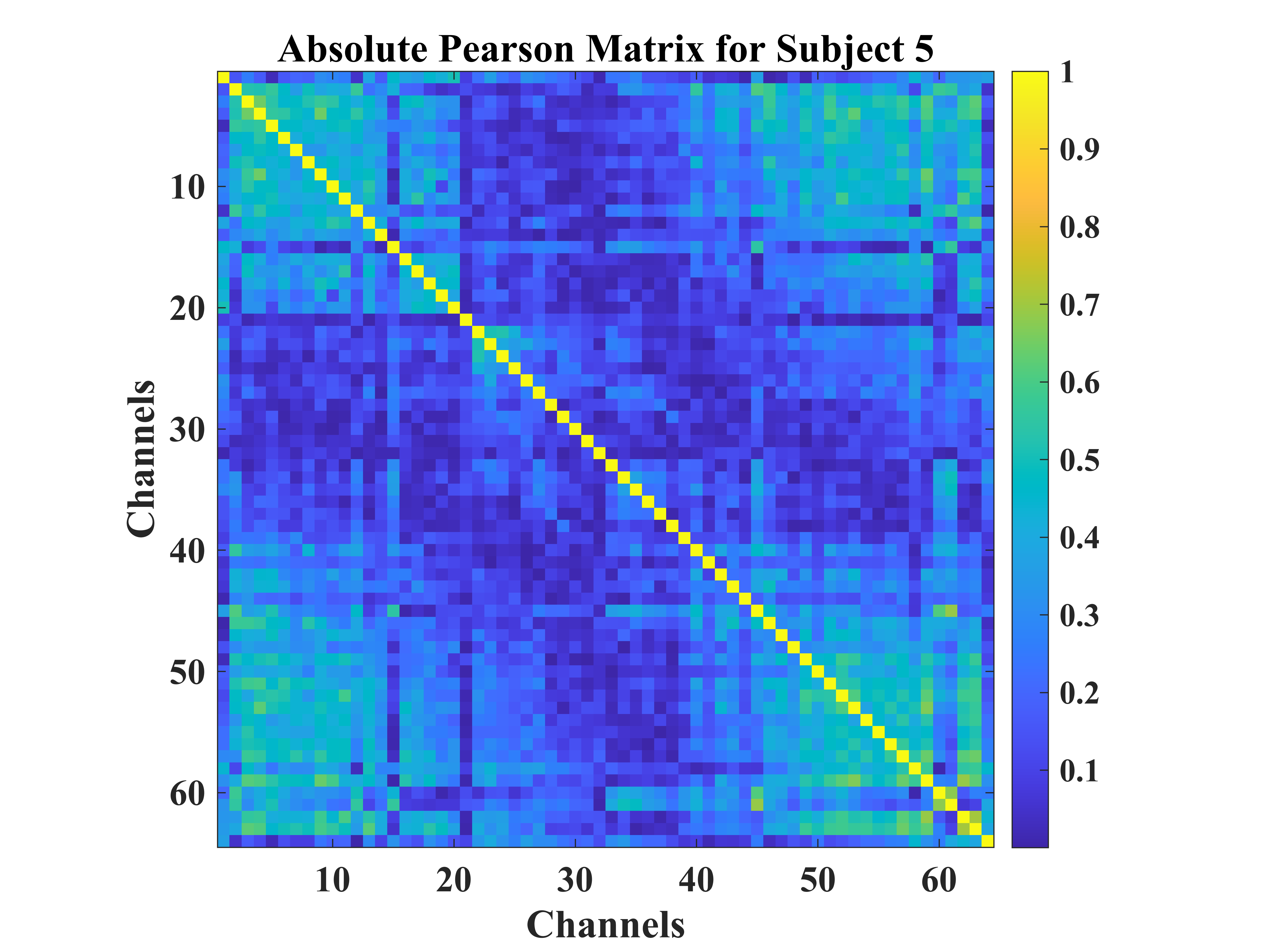}
			\subcaption{}
			\label{Absolute_Pearson_matrix_for_Subject_5}
		\end{minipage}
		\begin{minipage}[t]{.24\linewidth}
			\includegraphics[width=1.65in]{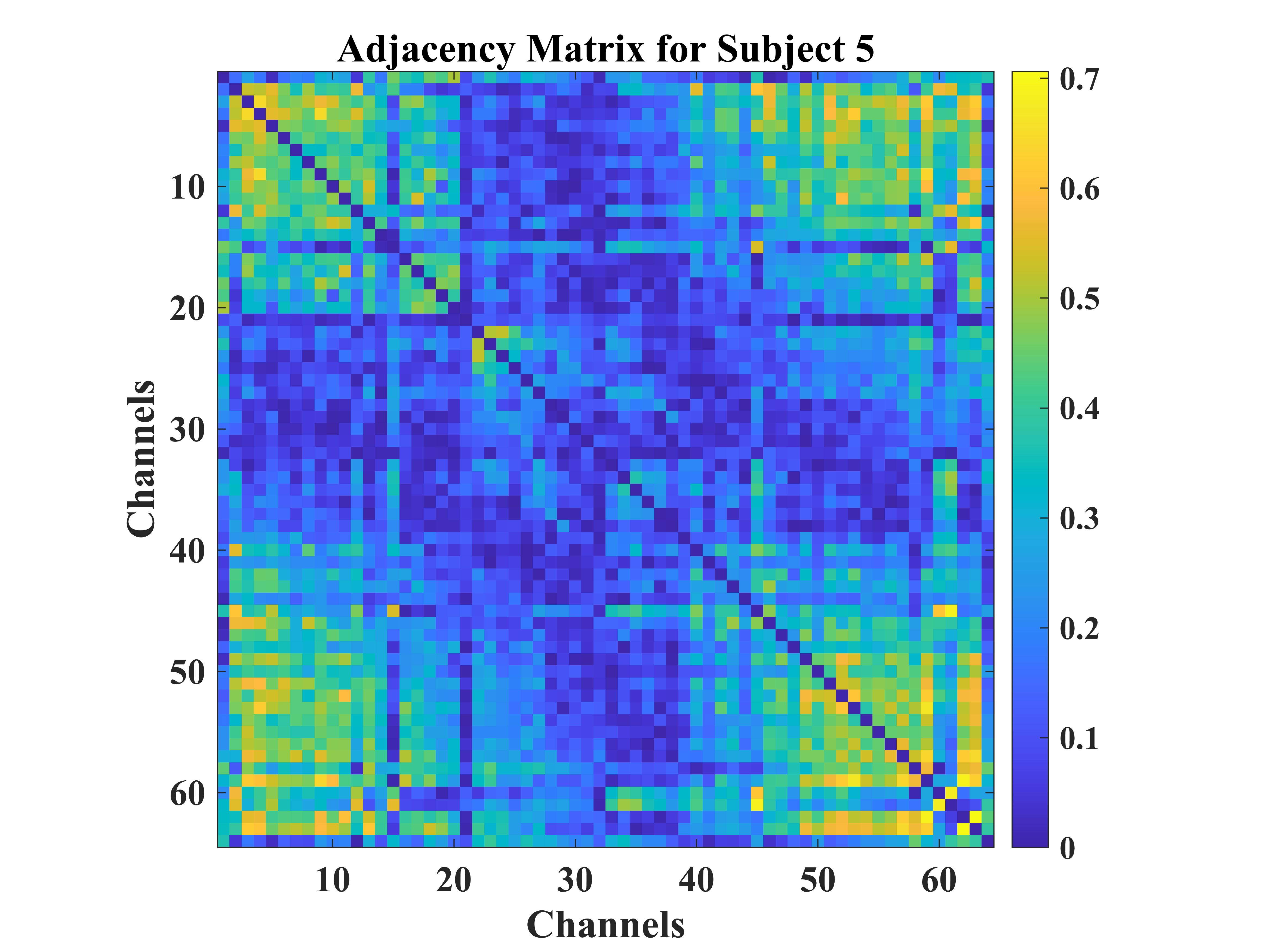}
			\subcaption{}
			\label{Adjacency_Matrix_for_Subject_5}
		\end{minipage}
		\begin{minipage}[t]{.24\linewidth}
			\includegraphics[width=1.65in]{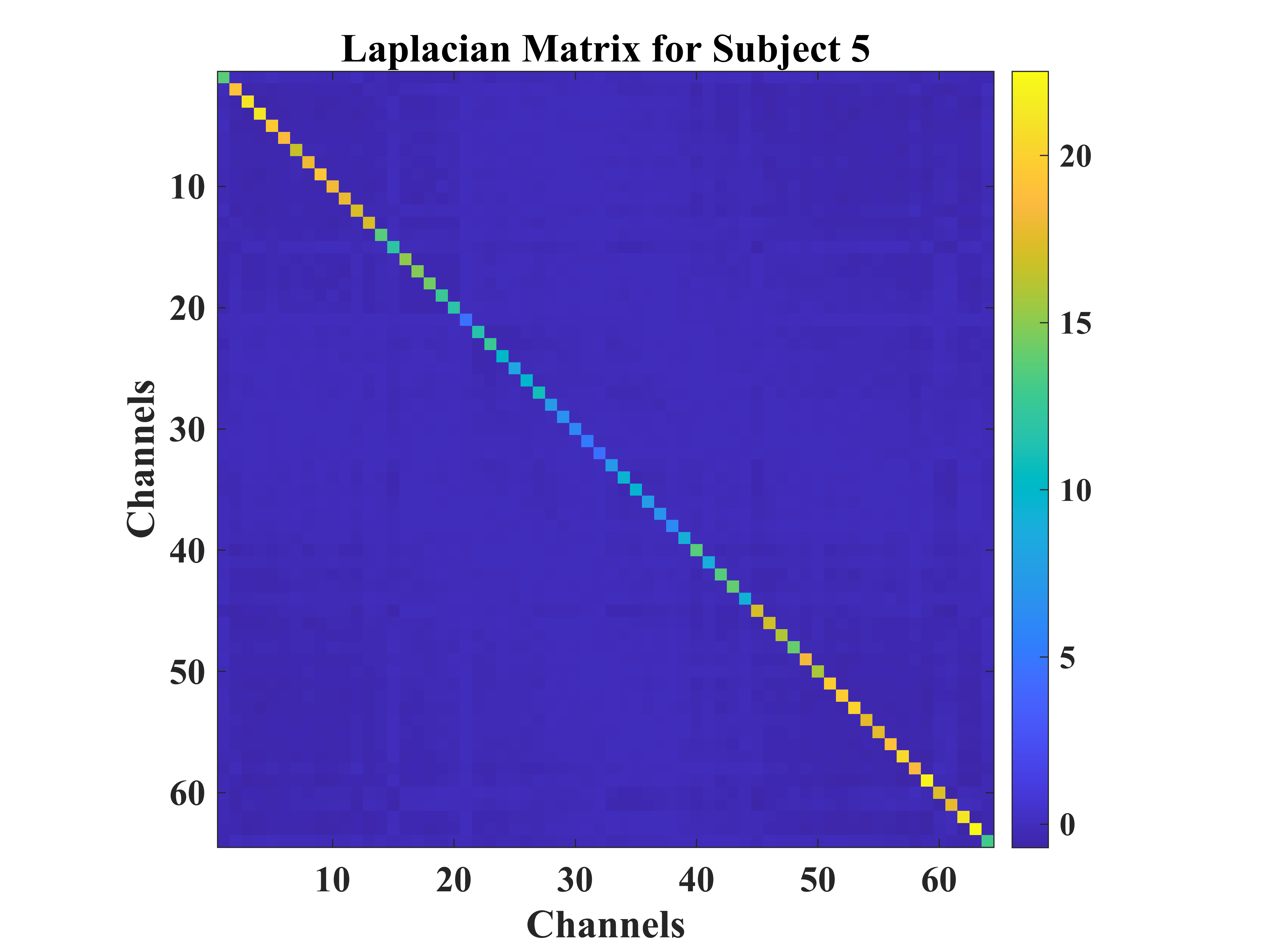}
			\subcaption{}
			\label{Laplacian_Matrix_for_Subject_5}
		\end{minipage}
		\caption{The PCC matrix, absolute PCC matrix, adjacency matrix, and graph Laplacian for Subject 10 and 5 from the PhysioNet Dataset. (a) The PCC matrix for Subject 10. (b) The absolute PCC matrix for Subject 10. (c) The adjacency matrix for Subject 10. (d) The graph Laplacian for Subject 10. (e) The PCC matrix for Subject 5. (f) The absolute PCC matrix for Subject 5. (g) The adjacency matrix for Subject 5. (h) The graph Laplacian for Subject 5.}
		\label{The PCC, Absolute PCC, adjacency and graph Laplacian matrices for Subject 10 and 5, separately}
	\end{figure*}
	
	In~\reffig{Global_Average_Accuracy_of_FC_Filters}, it shows that when there are more filters at the graph Conv layers, the accuracy ascends marginally. However, as indicated in~\reffig{loss of filters}, the loss value rises slightly after a fall. It means that the model with more filters has caused overfitting. The reason is that the model structure becomes much more complicated while applying more filters. Consequently, regarding the PhysioNet Dataset, 16, 32, 64, 128, 256, and 512 filters are used for the six-layer GCNs-Net to prevent overfitting. Meanwhile, there is a gentle descent of performances while adding more FC layers. As a result, a softmax layer is directly implemented without applying extra FC layers.
	
	\subsection{Subject-level Validation}\label{Subject-level Validation}
	The GCNs-Net is validated on 10 subjects from the PhysioNet Dataset, each with 64-channel 53,760 samples (640 time points $\times$ 84 trials $\times$ 1 subjects). The decoding accuracies are listed as follows: S$_1$ (97.08\%), S$_2$ (90.70\%), S$_3$ (97.92\%), S$_4$ (96.86\%), S$_5$ (80.49\%), S$_6$ (89.55\%), S$_7$ (84.82\%), S$_8$ (97.40\%), S$_9$ (97.02\%), and S$_{10}$ (98.72\%).
	
	According to~\reffig{The PCC, Absolute PCC, adjacency and graph Laplacian matrices for Subject 10 and 5, separately}, the PCC matrix, absolute PCC matrix, adjacency matrix, and graph Laplacian for Subject 10 and 5 are shown, which achieve the highest and the lowest accuracy. There are quite a lot of variations underlying the inter-subject EEG signals. For each one of the 10 subjects, 98.72\% maximum accuracy is achieved. As for the model of Subject 10, the AUC is 0.99. The single class accuracies on the L, R, B, and F are 99.92\%, 97.96\%, 98.08\%, and 98.93\%, respectively. For 10 subjects, the highest F1-score is 98.71\%, and the lowest is 80.19\%.
	
	Meanwhile, the presented GCNs-Net is validated on the High Gamma Dataset. The model containing 14 subjects is separately trained and evaluated. Since there are 44 electrodes~\cite{schirrmeister2017deep}, the maximum number of pooling layers is 2. We use the (C-C-C-P)$\times$2 architecture of the GCNs-Net to decode EEG tasks. From Subject 1 to Subject 14, 96.43\%, 95.63\%, 93.04\%, 99.18\%, 98.65\%, 94.77\%, 93.49\%, 97.91\%, 95.48\%, 96.77\%, 98.55\%, 98.69\%, 98.34\%, and 90.43\% accuracies are achieved, respectively. The mean accuracy is 96.24\%. The results indicate that the GCNs-Net manages to handle the individual variability due to its robustness and effectiveness.
	
	\subsection{Classification at the Group Level}\label{Classification Accuracy}
	Next, the GCNs-Net is evaluated at a group of 20 subjects (S$_1$$\sim$S$_{20}$) from the PhysioNet Dataset. The accuracy, Kappa, precision, recall, and F1-score are 88.35\%, 84.47\%, 88.39\%, 88.35\%, and 88.34\%, respectively. Further, the single class accuracies on the L, R, B, and F are 83.45\%, 86.72\%, 83.96\%, and 99.42\%, separately. The method performs well in classes F and B on the PhysioNet Dataset. The AUC is 0.92. Besides, it is evaluated on the High Gamma Dataset. The data of 14 subjects is used in the experiment. 80.89\% accuracy and 80.78\% F1-score are achieved. The reason for the accomplishment is that the GCNs-Net converges for the group-wise predictions, and succeeds in extracting relevant features from EEG signals.
	
	\subsection{10-fold Cross-validation for Reliability}\label{Repetitive Experiment}
	\begin{figure}[!ht]
		\centering
		\begin{minipage}[t]{.48\linewidth}
			\includegraphics[width=1.8in]{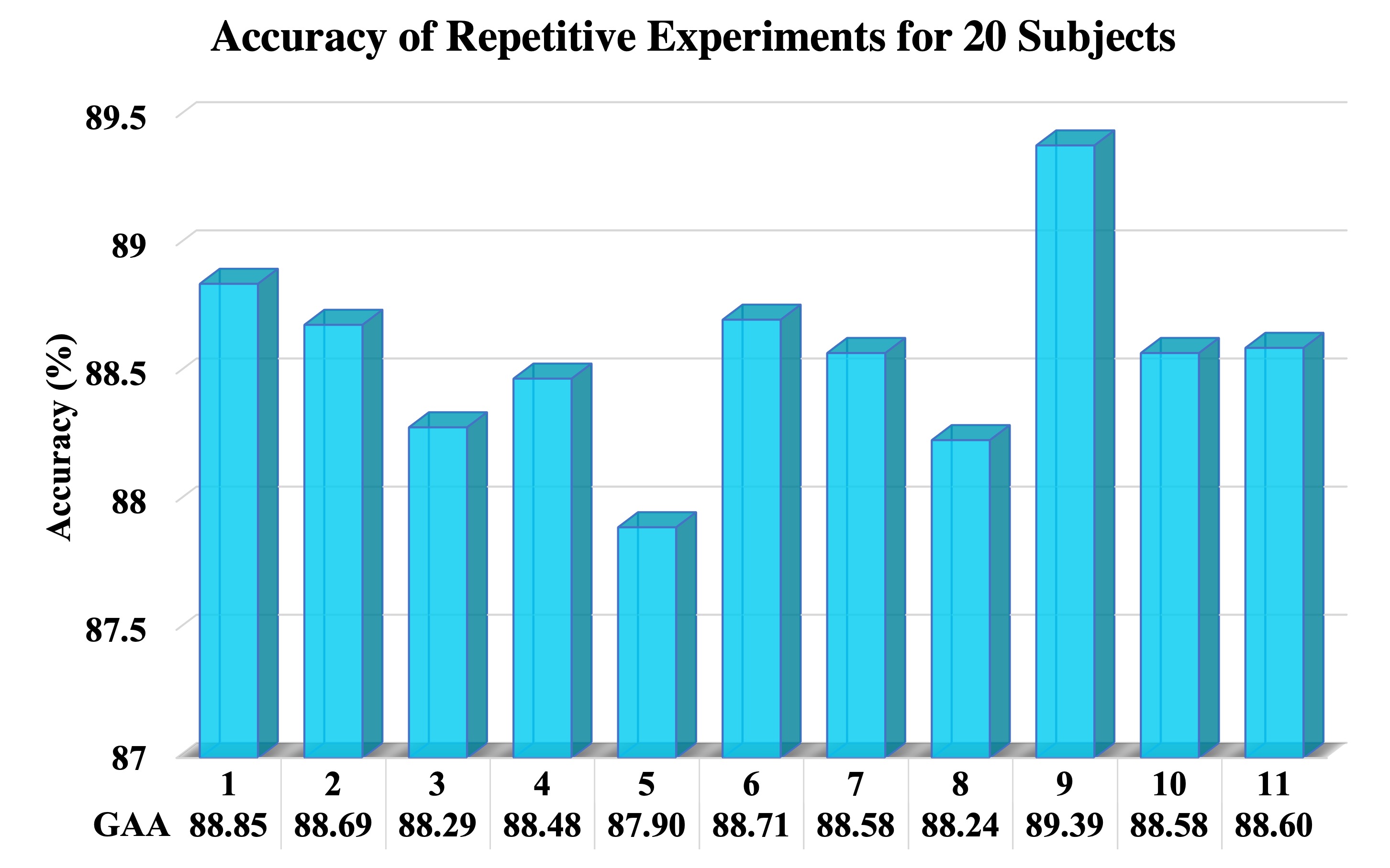}
			\subcaption{}
		\end{minipage}
		\begin{minipage}[t]{.48\linewidth}
			\includegraphics[width=1.8in]{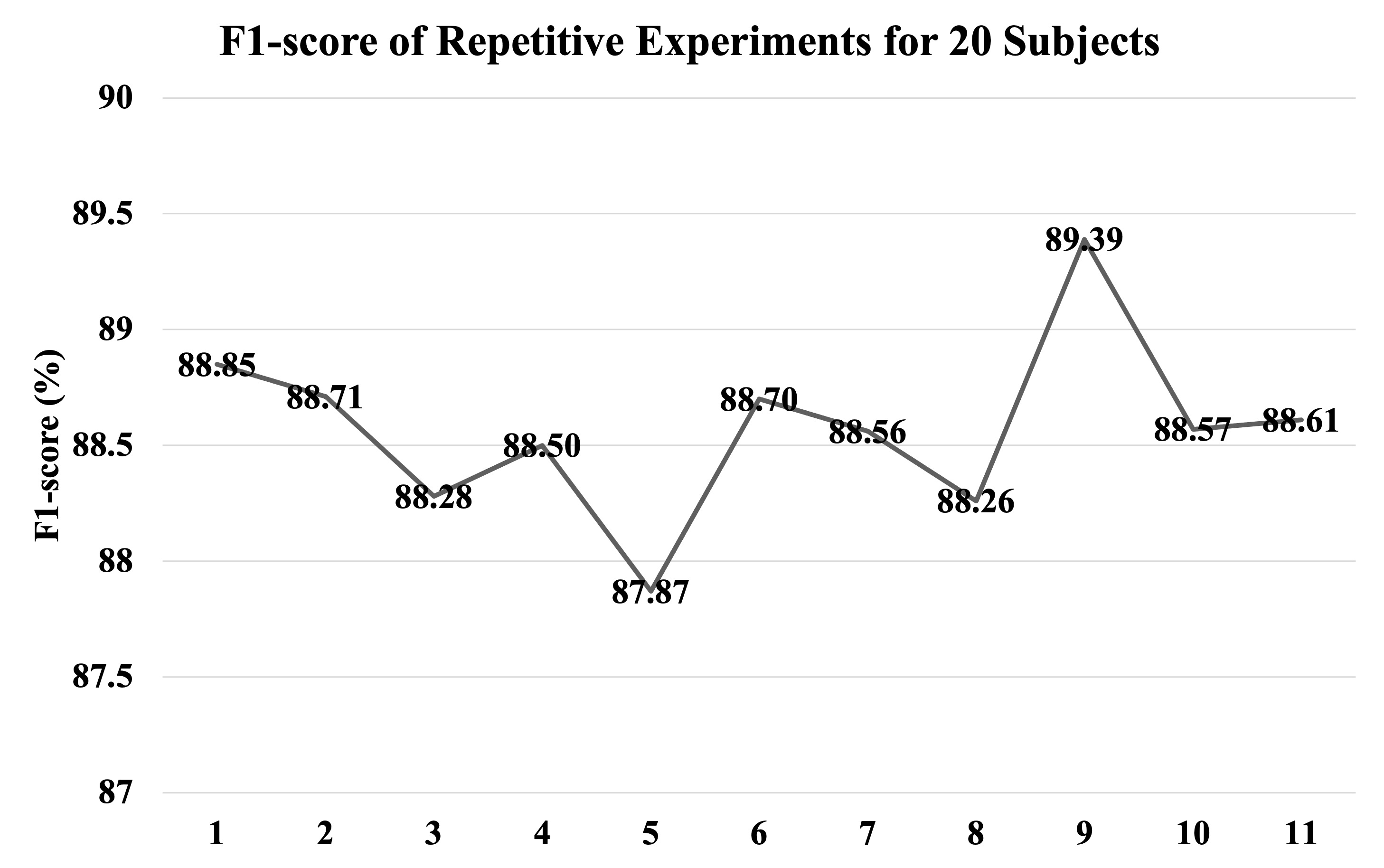}
			\subcaption{}
		\end{minipage}
		\caption{Accuracy and F1-score of the repetitive experiments for 20 subjects from the PhysioNet Dataset. (a) Accuracy of the repetitive experiments for 20 subjects. (b) F1-score of the repetitive experiments for 20 subjects.}
		\label{10-times}
	\end{figure}
	\begin{table*}[!ht]
		\centering
		\caption{Performance comparisons on the PhysioNet Dataset}
		\begin{tabular}{lcccccc}
			\toprule
			Related Work & Max. Accuracy & Avg. Accuracy & \emph{p}-value & Level & Approach & Num. of Subjects \\ \midrule
			
			\multirow{2}{*}{Dose~\etal (2018)~\cite{dose2018end}} & $-$ & 58.58\% & $-$ & Group & \multirow{2}{*}{CNNs} & 105 \\
			& 80.38\% & 68.51\% & $<0.05$ & Subject & & 1 \\
			
			Ma~\etal (2018)~\cite{ma2018improving} & 82.65\% & 68.20\% & $-$ & Group & RNNs & 12 \\ 
			
			\multirow{2}{*}{Hou~\etal (2020)~\cite{hou2019novel}} & 94.50\% & $-$ & $-$ & Group & \multirow{2}{*}{ESI-CNNs} & 10 \\
			& 96.00\% & $-$ & $>0.05$ & Subject & & 1 \\ 
			
			\multirow{2}{*}{Hou~\etal (2022)~\cite{hou2022deep}} & 94.64\% & $-$ & $-$ & Group & \multirow{2}{*}{BiLSTM-GCN} & 20 \\
			& 98.81\% & 95.48\% & $>0.05$ & Subject & & 1 \\
			
			\multirow{2}{*}{Jia~\etal (2022)~\cite{jia2020attention}} & 94.16\% & 93.78\% & $-$ & Group & \multirow{2}{*}{Graph ResNet} & 20 \\
			& 98.08\% & 94.18\% & $>0.05$ & Subject & & 1 \\
			
			\multirow{3}{*}{\textbf{Author}} & \textbf{89.39\%} & \textbf{88.57\%} & \multirow{3}{*}{$-$} & \multirow{2}{*}{\textbf{Group}} & \multirow{3}{*}{\textbf{GCNs-Net}} & \textbf{\begin{tabular}[c]{@{}c@{}}20\end{tabular}} \\ 
			& \textbf{88.14\%} & $-$ & & & & \textbf{100} \\ 
			& \textbf{98.72\%} & \textbf{93.06\%} & & \textbf{Subject} & & \textbf{1}\\
			\bottomrule
		\end{tabular}
		\label{Results Comparison-same dataset}
	\end{table*}
	
	The GCNs-Net is trained at the group level of 20 subjects (S$_1$$\sim$S$_{20}$) from the PhysioNet Dataset, following the 10-fold cross-validation to validate the stability and reliability. We divide the dataset into ten pieces and use one of them as the testing set, and the left nine pieces as the training set in turn.
	
	With the results from~\refsec{Novel Deep Learning Framework of the GCNs} (88.85\% accuracy, Model C6-P6-K2), 11 results are listed in~\reffig{10-times}. 89.39\% maximum accuracy is achieved, and the lowest is 87.90\%. The averaged accuracy and F1-score are both 88.57\%. At the group level, the performance is stably reproducible through repetitive experiments for cross-validation, showing the reliability and stability of the GCNs-Net.
	
	\subsection{Robustness to Data Size}\label{Robustness to Data Size}
\begin{figure}[h]
		\centering
		\begin{minipage}[t]{.48\linewidth}
			\includegraphics[width=1.8in]{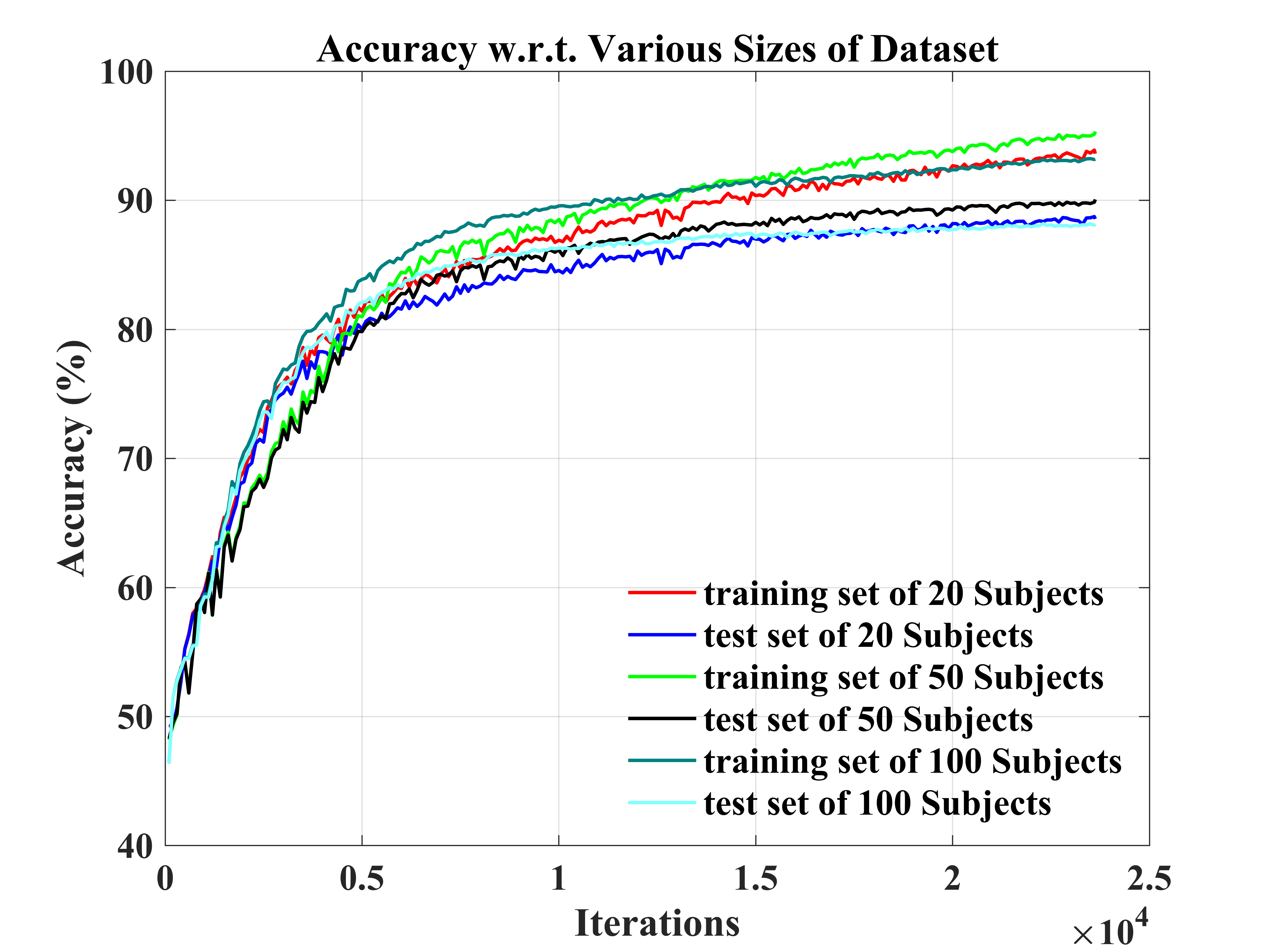}
			\subcaption{}
			\label{Accuracy Various Dataset}
		\end{minipage}
		\begin{minipage}[t]{.48\linewidth}
			\includegraphics[width=1.8in]{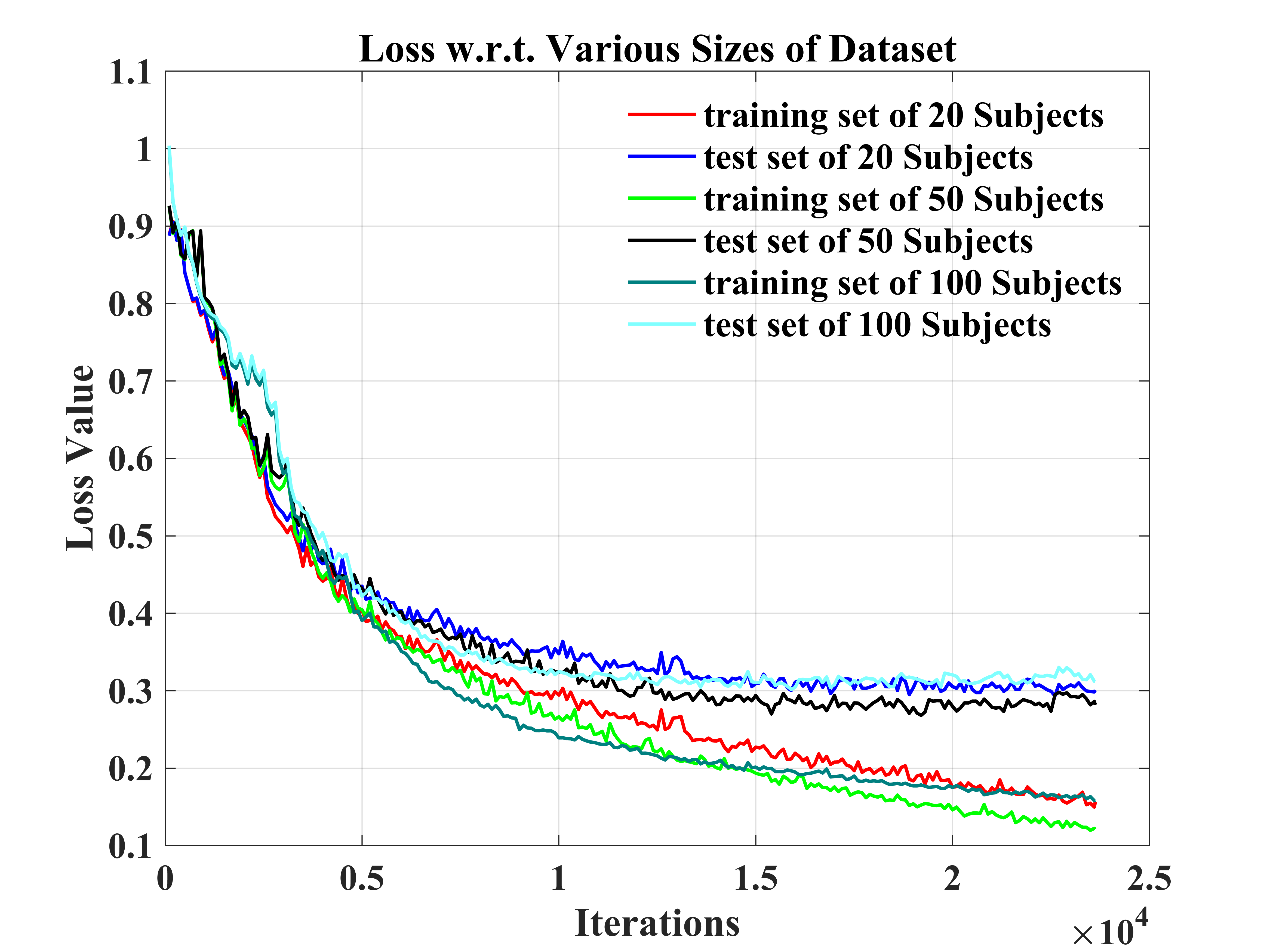}
			\subcaption{}
			\label{Loss Various Dataset}
		\end{minipage}
		\caption{Accuracy and loss regarding various sizes of datasets,~\ie, 20, 50, and 100 subjects, from the PhysioNet Dataset. (a) Accuracy regarding various sizes of datasets. (b) Loss regarding various sizes of datasets.}
		\label{Accuracy and Loss various dataset}
	\end{figure}

	It has also been trained and evaluated on different amounts of participants on the PhysioNet Dataset. The dataset of 50 subjects (S$_1$$\sim$S$_{50}$) with 64-channel 2,688,000 samples (640 time points $\times$ 84 trials $\times$ 50 subjects) and the dataset of 100 subjects (S$_1$$\sim$S$_{100}$) with 64-channel 5,376,000 samples (640 time points $\times$ 84 trials $\times$ 100 subjects) are used. Accuracies and losses are illustrated in~\reffig{Accuracy and Loss various dataset}.
	
	The GCNs-Net has achieved 89.75\% (testing set) and 94.99\% (training set) accuracies for a group of 50 subjects. Further, it also achieves 88.14\% (testing set) and 93.24\% (training set) accuracies for 100 subjects. The results have shown that it learns the generalized features from subjects at a larger scale. The reason for this phenomenon is that the GCNs-Net handles a larger amount of subjects, suggesting the adaptability and robustness to individual variability. The presented method is much more effective and robust in processing the graph-structured EEG-based MI signals since it has considered the functional topological relationship of EEG electrodes.
	
	\subsection{Comparison with State-of-the-art}\label{Performance Comparison}
	Two-level experiments are applied to compare the performance of the current models,~\ie, either from the subject level or the group level~\footnote{The reported results are derived from the corresponding papers. The ESI-CNNs (2020)~\cite{hou2019novel} method was re-implemented via the source codes provided by the authors: https://github.com/SuperBruceJia/EEG-Motor-Imagery-Classification-CNNs-TensorFlow.}. The decoding performance is mainly measured by the maximum accuracy (Max. Accuracy) and the averaged accuracy (Avg. Accuracy) on two datasets. First of all, the performance on the PhysioNet Dataset is compared in~\reftab{Results Comparison-same dataset}.
	
	The ESI-CNNs approach has achieved 94.50\% maximum accuracy at the group level (10 subjects)~\cite{hou2019novel}. Lately, the graph learning-based methods,~\eg, BiLSTM-GCN and Graph ResNet, achieve highly competitive performances. This phenomenon is due to the superiority of the graph representation learning for EEG signal processing~\cite{hou2022deep, jia2020attention}. The GCNs-Net also obtains competing performances, 89.39\% maximum accuracy for a group of 20 participants, and 88.14\% for 100 participants. Meanwhile, at the subject level, the $p$-value between the GCNs-Net and the CNNs model~\cite{dose2018end} is significantly less than 0.05. It indicates a significant difference in the predictive performance between two models, and the GCNs-Net is superior to predict EEG tasks, with a 30.21\% maximum accuracy increment. However, compared with the ESI-CNNs approach~\cite{hou2019novel}, there is no significant difference in classification performance at a 95\% confidence interval as the $p$-value is greater than 0.05. Furthermore, the GCNs-Net attains the best state-of-the-art performance at the hundred-subject level on the PhysioNet Dataset, which far exceeds current studies. The reason for the outcome is that the presented approach maintains robust and effective on the dataset with a larger amount of participants, regardless of the inter-trial and inter-subject variations.
	
	\begin{table}[h]
		\centering
		\footnotesize
		\caption{Performance comparisons on the High Gamma Dataset}
		\resizebox{\linewidth}{!}{
			\begin{tabular}{lccccc}
				\toprule
				Related Work & Avg. Accuracy & \emph{p}-value & Level & Approach & Dataset \\ \midrule
				Schirrmeister~\etal (2017)~\cite{schirrmeister2017deep} & 92.50\% & $<0.05$ & \multirow{3}{*}{Subject} & CNNs & \multirow{3}{*}{\begin{tabular}[c]{@{}c@{}}1 subjects\end{tabular}} \\
				Li~\etal (2019)~\cite{li2019channel} & 93.70\% & $<0.05$ & & CP-MixedNet & \\ 
				Tang~\etal (2020)~\cite{tang2020conditional} & 95.30\% & $>0.05$ & & DAN & \\ 
				\multirow{2}{*}{\textbf{Author}} & \textbf{80.89\%} & \multirow{2}{*}{$-$} & \textbf{Group} & \multirow{2}{*}{\textbf{GCNs-Net}} & \textbf{\begin{tabular}[c]{@{}c@{}}14 subject\end{tabular}} \\
				& \textbf{96.24\%} & & \textbf{Subject} & & \textbf{\begin{tabular}[c]{@{}c@{}}1 subject\end{tabular}} \\ 
				\bottomrule
			\end{tabular}
			\label{Results Comparison-Different dataset}
		}
	\end{table}
	
	In~\reftab{Results Comparison-Different dataset}, we compare the classification performance of some representative works~\cite{schirrmeister2017deep, li2019channel, tang2020conditional} on the High Gamma Dataset. Evaluated on the High Gamma Dataset, the $p$-values are both less than 0.05 while comparing the introduced method with the CNNs-based methods~\cite{schirrmeister2017deep, li2019channel}. The performance is significantly different among the models. The GCNs-Net successfully predicts MI tasks with dominant performances,~\ie, 99.18\% maximum accuracy, and 96.24\% averaged accuracy. The $p$-value compared with the DAN approach~\cite{tang2020conditional} is greater than 0.05. The performance of the two models is statistically less different, and both models achieve competing performances. Last but not least, the dominant classification accuracy has verified the robustness and effectiveness of our presented GCNs-Net.
	
	\section{Conclusion}\label{Conclusion}
	In order to deeply extract network patterns of brain dynamics, the GCNs-Net, a novel deep learning framework based on the GCNs, is presented to distinguish four-class MI intentions by cooperating with the functional topological relationship of EEG electrodes. The introduced method has been proven to converge for both personalized and group-wise predictions. Trained with individual data, the approach has achieved an averaged accuracy of 93.06\% (PhysioNet Dataset) and 96.24\% (High Gamma Dataset) in predicting the independent trials of the same participant, which is dominant in existing studies, indicating that the GCNs-Net converges well for individuals. Moreover, it has reached the uppermost accuracy on numerous sizes of group-level prediction on the PhysioNet Dataset,~\ie, with 89.39\% accuracy for 20 subjects, 89.75\% for 50 subjects, and 88.14\% for 100 subjects, which implies that it is considerably robust to individual variability. Further, it holds an averaged accuracy of 88.57\% after 10-fold cross-validation showing reliability and stability. On the other hand, it predicts all four MI tasks with superior accuracy, the best among which is the two feet prediction with an accuracy of 99.42\%. It indicates that the introduced method is able to build a generalized representation against both personalized and group-wise variations. In conclusion, we have developed a novel GCNs-Net method for EEG data classification. The outstanding performance of our method in MI identification is an important step towards better BCI approaches and neuroscience research.
	
	\section{Acknowledgment}\label{Acknowledgment}
	The authors appreciate the anonymous reviewers for their constructive comments to substantially improve this work. S. Jia personally would like to thank Prof. Y. Hou, Dr. J. Lv, Prof. Y. Li, Dr. Y. Shi, Prof. S. Zhang, and Prof. H. Yang for their patient guidance, encouragement, and helpful discussions, and this research paper would not have happened without them.
	
	\bibliographystyle{ieeetr}
	\bibliography{bibliography}
	
\end{document}